\documentclass[reprint, aps, prl, footinbib, letterpaper, superscriptaddress]{revtex4-2}

\usepackage[T1]{fontenc} 

\usepackage{amsmath,color}
\usepackage{amssymb}
\usepackage{amsfonts}
\usepackage{amsthm}
\usepackage{mathtools}

\usepackage{algorithm}
\usepackage{algpseudocode}
\usepackage{dsfont}

\usepackage{bbold}
\usepackage{chemformula}
\usepackage{siunitx}

\DeclarePairedDelimiter\bra{\langle}{\rvert}
\DeclarePairedDelimiter\ket{\lvert}{\rangle}
\DeclarePairedDelimiterX\braket[2]{\langle}{\rangle}{#1 \delimsize\vert #2}
\DeclarePairedDelimiterX\braket3[3]{\langle}{\rangle}{#1 \delimsize\vert #2 \delimsize\vert #3}

\usepackage{accents}

\usepackage{fancyvrb}

\newcommand{\hH}{\hat{H}}

\newcommand{\avg}[1]{\left\langle #1\right\rangle}

\newcommand{\red}[1]{{\color{black} #1}}

\begin{document}

	\title{Theory of Supervibronic Transitions via Casimir Polaritons}
	
	\author{Tao E. Li}%
	\email{taoeli@udel.edu}
	\affiliation{Department of Physics and Astronomy, University of Delaware, Newark, Delaware 19716, USA}
 
	\begin{abstract}
        A remote energy transfer pathway from electronic to vibrational degrees of freedom is identified inside an infrared optical cavity under vibrational strong coupling conditions. This mechanism relies on the dynamical Casimir effect, whereby real infrared photons are generated due to a sudden electronic transition of \red{anisotropic} molecules.  Moreover, the formation of vibrational polaritons enables the excited photon energy to be transferred to the vibrational degrees of freedom before any dissipation occurs. Both analytic solutions and numerical simulations reveal that the magnitude of this electronic to vibrational energy transfer depends quadratically on the number of molecules and resonantly on the vibration-cavity detuning. During this ``supervibronic'' transition process, because the vibrational energy gain per molecule can be meaningful in the macroscopic limit, this process may potentially be observed using conventional vibrational strong coupling devices.
	\end{abstract}

	\maketitle

    \paragraph{Introduction.} Exploring novel collective effects in light-matter interactions is a key research topic in the fields of quantum optics and chemical physics. \red{One early example is Dicke's superradiance \cite{Dicke1954,Gross1982,Breuer2007}, where the spontaneous emission rate of $N$ emitters interacting with a common electromagnetic field can be enhanced by a factor of $N$.  Beyond superradiance, when polaritons, the hybrid light-matter states, are formed under strong light-matter interactions, the Rabi splitting between a photon mode and $N$ molecular transitions can depend collectively on $\sqrt{N}$ \cite{Ribeiro2018,FriskKockum2019,Herrera2019,Nagarajan2021,Xiang2021JCP,Fregoni2022,Simpkins2023,Mandal2023ChemRev}. One popular experimental setup for preparing the polariton states is a planar Fabry--P\'erot microcavity, where  a cavity photon mode confined between a pair of parallel mirrors is tuned near resonance with a molecular transition inside the cavity.  More recently, under vibrational strong coupling (VSC)  between a molecular vibrational mode and an infrared (IR) cavity photon mode \cite{Shalabney2015,Long2015}, thermally-activated ground-state chemical reaction rates have been found to exhibit a nonlinear dependence on the concentration of molecules  in the Fabry--P\'erot cavity \cite{Thomas2016,Thomas2019_science,Thomas2020,Ahn2023Science}. Collective effects in polariton-induced vibrational or electronic energy transfer rates have also been observed experimentally \cite{Coles2014,Xiang2020Science,Grafton2020,Polak2020,Georgiou2021,Chen2022}. }

    A large amount of theoretical work has been reported in the past few years on the search of novel reaction or energy transfer pathways enabled by cavities \cite{Pino2015,Du2018,Saez-Blazquez2018,Galego2019,Semenov2019,Climent2019Plasmonic,Lacombe2019,Hoffmann2020,Li2020Origin,Perez-Sanchez2020,LiHuo2021,  Du2022, Mandal2020Polarized,Li2021Solute,Flick2017, Sidler2021,Yang2021QEDFT,Wei2021,TerryWeatherly2023,Li2023QMMM,Wei2023,Fischer2023,Sukharev2023,Aroeira2023}.  For one example, Mandel \textit{et al.} \cite{Mandal2020Polarized} numerically observed that, for a single \ch{LiH} model molecule coupled to a cavity mode at electronic transition frequencies, the permanent dipole moments of the molecule may generate photons during electronic excitations. This photon generation process is reminiscent of the dynamical Casimir effect \cite{Moore1970,Schwinger1993,Uhlmann2004,Dodonov2010,Macri2018}, a mechanism stating that real photons may be generated from the vacuum due to a fast change of boundary conditions or dielectric properties of the matter. 
    Because most strong coupling experiments are performed in the collective regime with the molecular number $N = 10^{6}\sim 10^{12}$, \cite{Ribeiro2018,Nagarajan2021,Simpkins2023} the transferability of these theoretical predictions made under single-molecule strong coupling to the collective regime remains unclear.

    \begin{figure}
		\centering
		\includegraphics[width=0.7\linewidth]{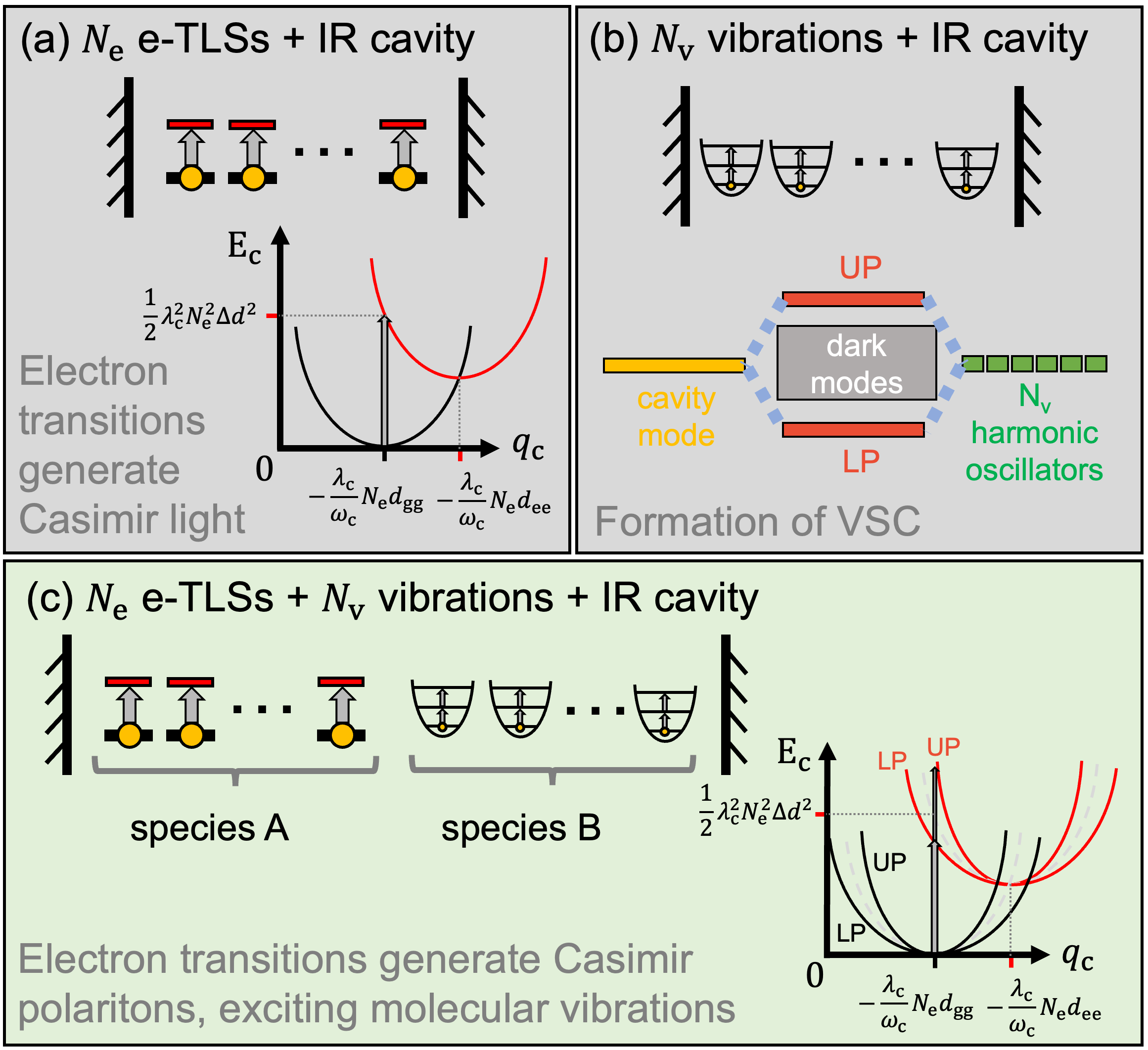}
	   \caption{The underlying mechanism of  ``supervibronic'' transitions \red{between two different species} (part c) can be decomposed to two more fundamental mechanisms: (a) dynamical Casimir effect whereby IR photons are generated due to the permanent dipole change during a sudden coherent electronic transition of $N_{\rm e}$ TLSs; and (b) collective VSC.} 
		\label{fig:demo}
    \end{figure}

    In this Letter, we theoretically report a remote energy transfer pathway from electronic to vibrational degrees of freedom enabled by collective VSC conditions. As illustrated in Fig. \ref{fig:demo}, this energy transfer pathway relies on the dynamical Casimir effect which generates IR cavity photons due to the permanent dipole change after a sudden electronic excitation of the molecules. Moreover, under collective VSC conditions, the excited IR photon mode can efficiently transfer energy to the vibrational degrees of freedom, thus achieving a remote electronic to vibrational energy transfer, or a remote vibronic transition. 
    
    \paragraph{Theory.} We start with the widely applied Pauli--Fierz Hamiltonian \cite{Flick2017,Hoffmann2020} for describing molecules coupled to a single-mode cavity:
    \begin{equation}\label{eq:H_PF_fundamental}
        \hH_{\rm PF} = \hH_{\rm M} + \frac{1}{2}\hat{p}_{\rm c}^2 + \frac{1}{2}\omega_{\rm c}^2 \left ( \hat{q}_{\rm c} + \frac{\boldsymbol{\lambda}_{\rm c}}{\omega_{\rm c}} \cdot \hat{\boldsymbol{\mu}} \right  )^2.
    \end{equation}
    Here, $\hH_{\rm M}$ denotes the standard molecular (kinetic + potential) Hamiltonian;  $\hat{p}_{\rm c}$, $\hat{q}_{\rm c}$, and $\omega_{\rm c}$ denote the momentum operator, position operator, and frequency of the cavity mode, respectively; $\boldsymbol{\lambda}_{\rm c}$ denotes the cavity coupling strength vector, which is oriented along the polarization direction of the cavity mode; $\hat{\boldsymbol{\mu}}$ denotes the total molecular dipole operator. 

    \red{We consider a molecular system} composed of  $N_{\rm e}$ electronic two-level systems (TLSs) plus $N_{\rm v}$ vibrational harmonic oscillators (Fig. \ref{fig:demo}c). The associated molecular Hamiltonian $\hH_{\rm M}$ can be written as
    \begin{equation}\label{eq:HM_model}
        \hH_{\rm M} = \omega_{\rm e} \sum_{i=1}^{N_{\rm e}} \hat{\sigma}_{+}^{i}\hat{\sigma}_{-}^{i} + \omega_{\rm v} \sum_{j=1}^{N_{\rm v}} \hat{b}_{j}^{\dagger}\hat{b}_{j} .
    \end{equation}
    In this non-interacting model Hamiltonian, both electronic and vibrational transition frequencies are uniform, equalling to $\omega_{\rm e}$ and $\omega_{\rm v}$, respectively. For each TLS indexed by $i$, the ground state and the excited state are labeled as $\ket{gi}$ and $\ket{ei}$, respectively;  $\hat{\sigma}_{-}^{i} = \ket{gi}\bra{ei}$, and $\hat{\sigma}_{+}^{i} = \ket{ei}\bra{gi}$. For each vibrational harmonic oscillator indexed by $j$, the creation and annihilation operators are denoted by $\hat{b}_{j}^{\dagger}$ and $\hat{b}_{j}$, respectively.

    The dipole moment directions of all electronic TLSs and vibrational harmonic oscillators are assumed to be aligned in parallel with the cavity polarization direction. With this important approximation in mind, below  $\boldsymbol{\lambda}_{\rm c}$ and $\hat{\boldsymbol{\mu}}$  in Eq. \eqref{eq:H_PF_fundamental} will be replaced by the corresponding scalar quantities $\lambda_{\rm c}$ and $\hat{\mu}$, respectively. \red{$\hat{\mu}$ is represented as the sum of the electronic and vibrational components}: $\hat{\mu} = \hat{\mu}_{\rm e} + \hat{\mu}_{\rm v}$. The electronic dipole operator is \cite{Mandal2020Polarized}
    \begin{equation}\label{eq:dipole_e_operator}
        \hat{\mu}_{\rm e} = \sum_{i=1}^{N_{\rm e}} d_{\rm eg}( \hat{\sigma}_{+}^{i} + \hat{\sigma}_{-}^{i} ) + \sum_{i=1}^{N_{\rm e}}\left( \bar{d} \hat{1}^i - \frac{\Delta d}{2} \hat{\sigma}_z^{i} \right).
    \end{equation}
    Here, $\hat{1}^i = \ket{gi}\bra{gi} + \ket{ei}\bra{ei}$;  $\hat{\sigma}_z^{i} = \ket{gi}\bra{gi} - \ket{ei}\bra{ei}$; 
    $d_{\rm eg}$ denotes the transition dipole moment;
    $\bar{d} \equiv (d_{\rm gg} + d_{\rm ee}) / 2$ and $\Delta d  \equiv d_{\rm ee} - d_{\rm gg}$, where $d_{\rm gg}$ and $d_{\rm ee}$ denote the permanent dipole moments of the ground and the excited state, respectively. 
    For the vibrational harmonic oscillators, the dipole  operator $\hat{\mu}_{\rm v}$ is
    \begin{equation}\label{eq:mu_v_qm}
        \hat{\mu}_{\rm v} = \sum_{j=1}^{N_{\rm v}} \frac{1}{\sqrt{2\omega_{\rm v}}} d_{\rm v} \left( \hat{b}^{\dagger}_{j} + \hat{b}_j \right),
    \end{equation}
    where $d_{\rm v}$ denotes the transition dipole moment for each vibrational harmonic oscillator.

    \paragraph{Analytic solution.} For our problem, because the photonic and vibrational degrees of freedom have similar transition frequencies, we may generalize the Ehrenfest approximation used in nonadiabatic molecular dynamics \cite{Li2005Eh} and treat both the photonic and vibrational dynamics classically, leaving only the electronic dynamics quantum-mechanically.
    With this cavity Ehrenfest approximation \cite{Miller1978,Li2018Spontaneous,Hoffmann2018} in mind, now, let us consider a scenario in which the coupled electron-vibration-cavity system starts from the global ground state and then a delta pulse suddenly excites the electronic TLSs. Under this case, due to the timescale separation, the photonic and vibrational degrees of freedom can be assumed to be frozen during the electronic transition, an approximation which may be called as the cavity Frank--Condon transition. Then, the change of photonic energy \cite{Schafer2020,Taylor2020,Stokes2022,Mandal2023ChemRev} during this electronic transition, according to Eq. \eqref{eq:H_PF_fundamental}, can be written as 
    $\Delta E_{\rm c} = \frac{1}{2}\omega_{\rm c}^2\left (q_{\rm c} + \frac{\lambda_{\rm c}}{\omega_{\rm c}}\avg{\hat{\mu}}_{t_{0^+}} \right )^2 - \frac{1}{2}\omega_{\rm c}^2\left (q_{\rm c} + \frac{\lambda_{\rm c}}{\omega_{\rm c}}\avg{\hat{\mu}}_{t_{0^-}} \right )^2$ \footnote{The classical photonic energy is calculated by $\frac{1}{2}p_{\rm c}^2 + \frac{1}{2}\omega_{\rm c}^2(q_{\rm c} + \lambda_{\rm c} \avg{\hat{\mu}}/\omega_{\rm c} )^2$ instead of $\frac{1}{2}p_{\rm c}^2 + \omega_{\rm c}\lambda_{\rm c} q_{\rm c}\avg{\hat{\mu}} + \frac{1}{2}\lambda_{\rm c}^2 \avg{\hat{\mu}^2}$ due to a recent suggestion \cite{Taylor2020} considering gauge invariance.},
    where $\avg{\hat{\mu}}_{t_{0^-}}$  and $\avg{\hat{\mu}}_{t_{0^+}}$ denote the mean-field average of the total molecular dipole moments before and after the electronic transition at time $t_0$, respectively. Due to the use of the cavity Frank--Condon approximation, the cavity photon position $q_{\rm c}$ is assumed fixed during the electronic transition process, the value of which is $q_{\rm c} = -\frac{\lambda_{\rm c}}{\omega_{\rm c}} \avg{\hat{\mu}}_{t_{0^-}}$. This is because at the global ground state (before the electronic excitation), the minimization of total energy requires the cavity photon position to be displaced by the molecular polarization \cite{Schafer2020,Mandal2020Polarized}.  Substituting the value of $q_{\rm c}$ into the form of $\Delta E_{\rm c}$, we find 
    \begin{equation}\label{eq:E_ph_change}
        \Delta E_{\rm c} = \frac{1}{2}\lambda_{\rm c}^2(\avg{\hat{\mu}}_{t_{0^+}} - \avg{\hat{\mu}}_{t_{0^-}})^2 .
    \end{equation}
    Say, if the electronic density matrix per TLS changes from $\bigl( \begin{smallmatrix}1 & 0\\ 0 & 0\end{smallmatrix}\bigr)$ to $\bigl( \begin{smallmatrix}1 - P_{\rm e} & \rho_{\rm ge}\\ {\rho_{\rm eg}} & P_{\rm e}\end{smallmatrix}\bigr)$ during the electronic transition, using Eq. \eqref{eq:dipole_e_operator}, we may write $\avg{\hat{\mu}}_{t_{0+}} - \avg{\hat{\mu}}_{t_{0-}} = N_{\rm e}(P_{\rm e}\Delta d + 2d_{\rm eg} \Re\rho_{\rm eg})$, where $\Re\rho_{\rm eg}$ denotes the real component of the off-diagnoal electronic coherence.  Therefore, the final change of photonic energy becomes
    \begin{equation}\label{eq:E_ph_change_final}
        \Delta E_{\rm c} = \frac{1}{2}\lambda_{\rm c}^2 N_{\rm e}^2 \left ( P_{\rm e}\Delta d + 2d_{\rm eg} \Re\rho_{\rm eg} \right )^2.
    \end{equation}
    Because the molecular number is  large under collective strong coupling, it becomes hard to maintain a significant electronic coherence among all TLSs over long times. Therefore, one may treat  $\Re\rho_{\rm eg} \rightarrow 0$ as the long-time limit and obtain a simpler form  of the photonic energy change: 
    \begin{equation}\label{eq:E_ph_change_final2}
        \Delta E_{\rm c} \rightarrow \frac{1}{2}\lambda_{\rm c}^2 N_{\rm e}^2 P_{\rm e}^2\Delta d^2.
    \end{equation}
    Eq. \eqref{eq:E_ph_change_final2} suggests that a sudden change in the permanent dipole moment \red{($\Delta d$)} may generate real photons in IR cavities. \red{Our simulations below show that $\Delta E_{\rm c}/\omega_{\rm c} \gg 1$, indicating the generation of a substantial number of IR photon quanta. This  photon generation process can be interpreted as the release of the dipole self-energy contribution ($\frac{1}{2}\lambda_{\rm c}^2\avg{\hat{\mu}}^2$) in the IR photon energy via sudden electronic transitions.}  Because the permanent dipole moment determines the dielectric properties, this finding aligns with previous \red{theoretical} work on the dynamical Casimir effect \cite{Uhlmann2004,Dodonov2010} \red{discussing} generating real photons via fast changes of the dielectric properties of the material\red{, which requires further experimental verification.} A more intuitive understanding of \red{our Casimir photon generation process} is shown in Fig. \ref{fig:demo}a. 
    

    Under VSC, because the IR photon mode is hybridized with the vibrational bright mode, the sudden electronic transition can directly excite vibrational polaritons with the dynamical Casimir effect. These excited polaritons can be called as Casimir polaritons. At resonance conditions (i.e., when $\omega_{\rm v} = \omega_{\rm c}$), 
    because the polaritons are an equal mixture of photonic and vibrational components (Fig. \ref{fig:demo}b),  
    the vibrational degrees of freedom may receive a total amount of energy $\Delta E_{\rm v} \approx \Delta E_{\rm c} / 2$ \footnote{Here no dissipation pathway is considered}. When the photonic frequency is very different from the vibrational frequency, vibrational polaritons are no longer formed, so $\Delta E_{\rm v} \rightarrow 0$. With these two limits in mind, we can approximately write  the energy gain in the vibrational degrees of freedom as
    \begin{equation}\label{eq:Delta_E_v}
        \Delta E_{\rm v} \approx \frac{1}{2}\Delta E_{\rm c}\rho(\omega_{\rm v} - \omega_{\rm c}),
    \end{equation}
    where the density of states $\rho(\omega_{\rm v} - \omega_{\rm c}) = 1$ at resonance and $\rho(\omega_{\rm v} - \omega_{\rm c}) \rightarrow 0$ when $|\omega_{\rm v} - \omega_{\rm c}|$ becomes large. Per vibrational degree of freedom, the energy gain is $\Delta E_{\rm v} /N_{\rm v} \approx \frac{1}{2 N_{\rm v}}\Delta E_{\rm c}\rho(\omega_{\rm v} - \omega_{\rm c}) \red{\ \propto N_{\rm e}^2/N_{\rm v}}$. 
    If $N_{\rm e} \sim N_{\rm v}$, 
     each vibrational harmonic oscillator may receive an energy gain proportionally to the total molecular number in the cavity.  This analysis  indicates the existence of a remote, collective transition from the electronic to the vibrational degrees of freedom bridged by the IR cavity mode (or Casimir polaritons). An intuitive understanding of this dynamical-Casimir-effect-induced vibronic transition is given in the energy diagram of Fig. \ref{fig:demo}c. Here, under an external pulse excitation, the IR photon mode experiences a vertical transition from the vibrational polariton states dressed by the electronic ground state (black curves) to the vibrational polariton states dressed by the electronic excited state (red curves), thus driving the oscillations of molecular vibrations. For want of a better term, this collective vibronic transition process can be called as ``supervibronic'' transitions.

    \begin{figure}
		\centering
		\includegraphics[width=0.8\linewidth]{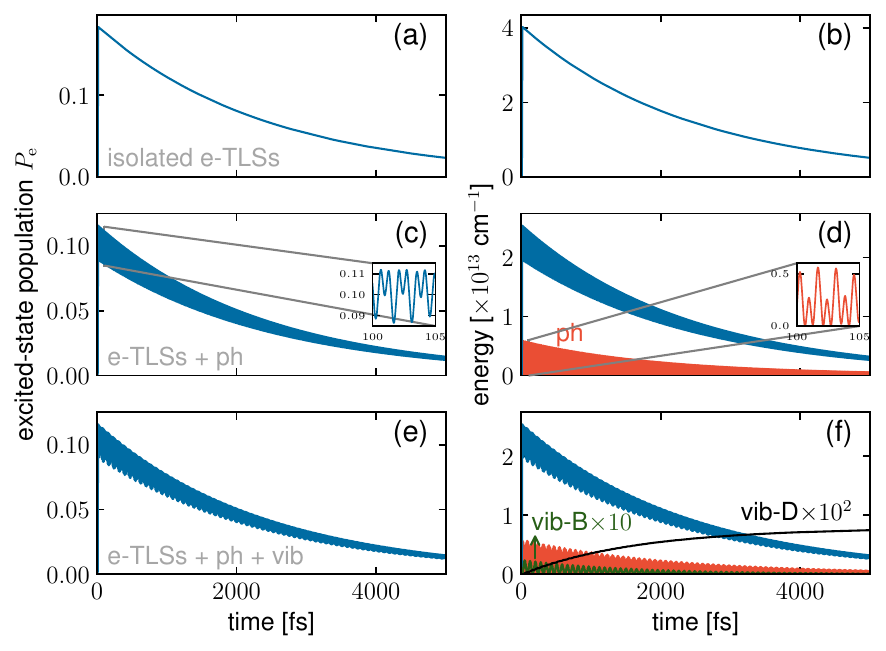}
		\caption{Electronic excited-state population $P_{\rm e}$ dynamics (left panel) and energy dynamics of each component (right panel) during a Gaussian pulse excitation under different scenarios: (a,b) $N_{\rm e}$ isolated electronic TLSs; (c,d) $N_{\rm e}$ electronic TLSs coupled to an IR cavity mode; (e,f) $N_{\rm e}$ electronic TLSs plus $N_{\rm v}$ vibrational harmonic oscillators coupled to an IR cavity mode. Lines of different colors are used to represent the energy of different components: electronic TLSs (blue); cavity mode (red); bright mode of vibrations (green); dark modes of vibrations (black). For better visualization, the energy of the bright and the dark modes is amplified by a factor of 10 and $10^2$, respectively. 
		}
		\label{fig:elem_processes}
    \end{figure}

    \paragraph{Simulation results.} To better investigate the possibility of  supervibronic transitions under realistic experimental conditions, we now perform numerical investigation by including three possible relaxation channels of the system: electronic depopulation and decoherence, cavity loss, and polariton dephasing to the dark modes \cite{Xiang2019State,Li2020Nonlinear}. \red{Due to the symmetry of the system, the electronic subsystem is represented by the density matrix of a single TLS $\hat{\rho}_{\rm e,s}$, an assumption which is justified by comparing to the results from an explicit simulation of $N_{\rm e}$ TLSs when $N_{\rm e}$ is small. The dynamics of $\hat{\rho}_{\rm e,s}$ is  governed by the mean-field Hamiltonian: $\hH_{\rm sc,s} = \omega_{\rm e} \hat{\sigma}_{+} \hat{\sigma}_{-}  + \omega_{\rm c}\lambda_{\rm c}q_{\rm c} \hat{\mu}_{\rm e,s} + \frac{1}{2}\lambda_{\rm c}^2 \left [\hat{\mu}_{\rm e,s}^2 + 2(N_{\rm e}-1)\avg{\hat{\mu}_{\rm e,s}}\hat{\mu}_{\rm e,s} + 2\mu_{\rm v}\hat{\mu}_{\rm e,s}\right ]$, where $\hat{\mu}_{\rm e,s}$ and $\mu_{\rm v}$ denote the single-body electronic dipole operator and the total vibrational dipole moment, respectively.   The vibrational dynamics are explicitly propagated in a basis of vibrational bright and dark modes.} See the SI for details on numerical simulations. 
    
    To start with, we analyze the time-dependent dynamics for an elementary process: $N_{\rm e}$ isolated electronic TLSs under a Gaussian pulse excitation. This Gaussian pulse is applied at $t = 12.1$ fs with a width of 2.4 fs and an amplitude of $E_0=0.01$ a.u. The role of this pulse is to induce a sudden electronic transition. Fig. \ref{fig:elem_processes}a,b plot the corresponding excited-state population ($P_{\rm e}$) dynamics per TLS and the total energy dynamics of all the TLSs, respectively. Not surprisingly, due to the electronic depopulation and decoherence, both signals experience an exponential decay with a lifetime $\tau_{\rm e} \sim 2.4$ ps after the pulse pumping. 
    
    Then, we study the case when the electronic TLSs are coupled to the IR cavity mode (Fig. \ref{fig:demo}a). After an initial Gaussian pulse excitation, the electronic excited-state population $P_{\rm e}$ and the total electronic energy experience an exponential decay modulated by fast oscillations, as shown in Figs. \ref{fig:elem_processes}c,d. At the same time, the photonic energy (red line in Fig. \ref{fig:elem_processes}d), as calculated by $\frac{1}{2}p_{\rm c}^2 + \frac{1}{2}\omega_{\rm c}^2(q_{\rm c} + \lambda_{\rm c} \avg{\hat{\mu}} / \omega_{\rm c})^2$, is also excited and modulated by fast oscillations. The similar fast-oscillation patterns in electronic and photonic energy dynamics \red{(see the insets)} can be understood from Eq. \eqref{eq:E_ph_change_final}. This equation implies that the photonic energy change is determined by the $\Re\rho_{\rm eg}$ term, 
    which exhibits fast oscillations. In later times, due to the electronic relaxation and decoherence as well as the cavity loss, both the electronic and photonic energies are quenched. Overall, Figs. \ref{fig:elem_processes}c,d confirms the dynamical Casimir effect that the IR photon mode can be transiently excited during a sudden electronic excitation of the TLSs.

    After studying the above two processes, we  study the case of our interest --- both $N_{\rm e}$ electronic TLSs and $N_{\rm v}$ harmonic oscillators are coupled to the IR cavity mode.
    Fig. \ref{fig:elem_processes}e plots the dynamics of the electronic excited-state population $P_{\rm e}$ after an initial Gaussian pulse excitation of the TLSs. This pattern is similar to the case of the coupled electron-cavity system in Fig. \ref{fig:elem_processes}c. As far as the energy dynamics of different subsystems are considered, while the electronic and photonic responses are also similar to the case of the coupled electron-cavity system (Fig. \ref{fig:elem_processes}d), the vibrational bright mode (green line) is also initially excited during the electronic excitation. This is because, at resonance conditions, the cavity mode forms polaritons with the vibrational bright mode. Thus, the IR mode excited due to the dynamical Casimir effect coherently transfers its energy to the vibrational bright mode before any dissipation occurs. The signature of vibrational polaritons can be found in the oscillation pattern of the bright mode, which exhibits a period of $\sim 78$ fs, corresponding to the Rabi splitting of 430 cm$^{-1}$. 
    
    As shown in Fig. \ref{fig:elem_processes}f, during the relaxation of the electronic, photonic, and bright-mode signals, the vibrational dark modes (black line) gradually gain energy due to the polariton dephasing mechanism. At $t = 5$ ps, the energy gain of the vibrational dark modes reaches $7.6\times 10^{10}$ cm$^{-1}$. Because $N_{\rm v} = 10^{10}$ is used during the simulation (see the SI for details), the energy gain per vibrational harmonic oscillator is $7.6$ cm$^{-1}$ (= 11 K). 
    This result numerically validates the existence of a remote energy transfer pathway from the electronic to the vibrational degrees of freedom, in spite of the inclusion of three dissipation channels.

    \begin{figure}
		\centering
		\includegraphics[width=1.0\linewidth]{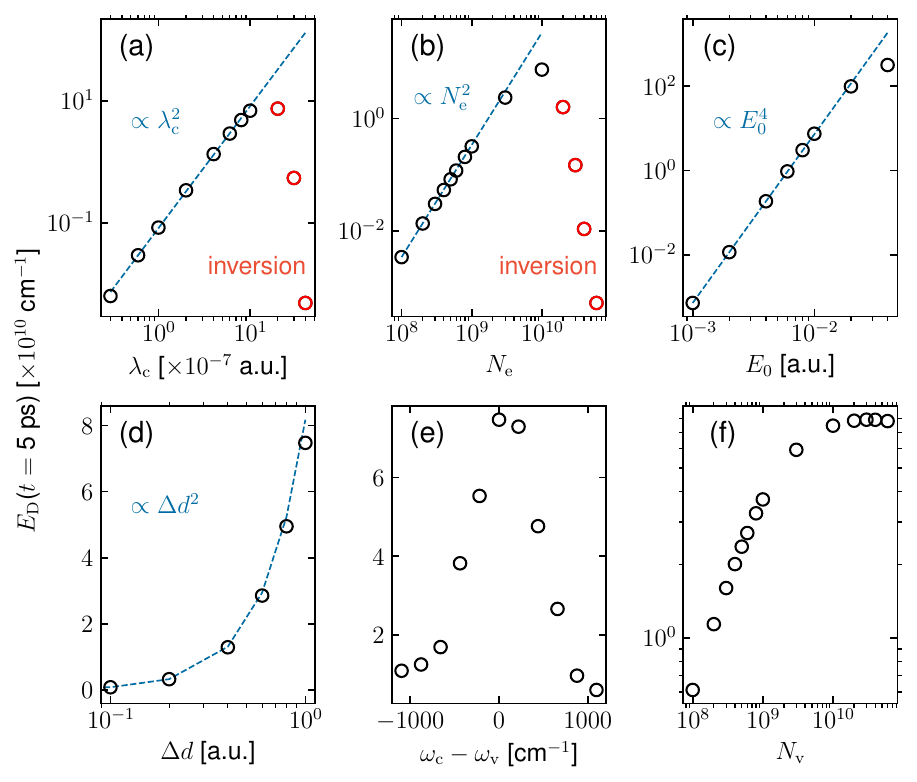}
		\caption{
  Transient energy of the vibrational dark modes at $t = 5$ ps [$E_{\rm D}(t = 5 \text{ ps})$] against different parameters in the logarithmic scale: (a) $\lambda_{\rm c}$; (b) $N_{\rm e}$; (c) $E_0$; (d) $\Delta d$; (e)  $\omega_{\rm c} - \omega_{\rm v}$; (f) $N_{\rm v}$. Open circles are the simulation data, and blue dashed lines are the corresponding asymptotes defined by Eqs.  \eqref{eq:E_ph_change_final2} and \eqref{eq:Delta_E_v}. 
		}
		\label{fig:params_dependence}
    \end{figure}

    In Figs. \ref{fig:params_dependence}, we perform additional numerical simulations to examine the analytic solution of the long-time vibrational energy gain ($\Delta E_{\rm  v}$) in Eqs. \eqref{eq:E_ph_change_final2} and \eqref{eq:Delta_E_v}, using the energy gain of vibrational dark modes at $t = 5$ ps [$E_{\rm D}(t = 5 \text{ ps})$] to measure the long-time $\Delta E_{\rm v}$.
    Figs. \ref{fig:params_dependence}a-e plot  $E_{\rm D}(t = 5 \text{ ps})$  against the cavity-matter coupling ($\lambda_{\rm c}$), the number of electronic TLSs ($N_{\rm e}$), the external driving field amplitude ($E_0$), the permanent dipole difference between the electronic ground and the excited state ($\Delta d$), and the vibration-cavity detuning ($\omega_{\rm v} - \omega_{\rm c}$). Overall, the parameter dependence agrees well with Eqs. \eqref{eq:E_ph_change_final2} and \eqref{eq:Delta_E_v}, exhibiting a scaling depending quadratically on $\lambda_{\rm c}$, $N_{\rm e}$, and $\Delta d$, fourth-order on $E_0$, in addition to a resonant dependence on $\omega_{\rm v} - \omega_{\rm c}$. Here, the $E_0^4$ dependence is equivalent to a quadratic dependence on $P_{\rm e}$ in Eq. \eqref{eq:E_ph_change_final2}.
    This is because, before saturation,  $P_{\rm e}$ depends linearly against the pulse power, and the pulse power is proportional to $E_0^2$. The parameter dependence against the number of vibrational harmonic oscillators $N_{\rm v}$ is also plotted in Fig. \ref{fig:params_dependence}f, showing a null dependence against $N_{\rm v}$ when $N_{\rm v}$ is sufficiently large, in agreement with Eqs. \eqref{eq:E_ph_change_final2} and \eqref{eq:Delta_E_v}.
    
    However, in Figs. \ref{fig:params_dependence}, the only discrepancy with the analytic solution occurs at the limit when $\lambda_{\rm c}$ or $N_{\rm e}$ becomes very large (the red circles).  In these situations, the simulation data show that $E_{\rm D}(t = 5 \text{ ps})$ decreases when $\lambda_{\rm c}$ or $N_{\rm e}$ further increases. \red{These inversion regimes arise because the pulse excitation of the electronic population ($P_{\rm e}$) becomes less efficient  as the electron-cavity coupling strength (which depends on either $\lambda_{\rm c}$ or $N_{\rm e}$) increases due to a deviation from the resonant excitation condition, an effect not accounted for in the derivation of Eqs. \eqref{eq:E_ph_change_final2} and \eqref{eq:Delta_E_v}.}
    See also the SI for additional discussions on the inversion regimes. Fig. S2 in the SI also demonstrates that this supervibronic transition pathway is robust under a very wide range of the dissipative rates.

    \red{Our calculations above assume that the orientations of both the electronic and the vibrational dipole moments  align in parallel with the cavity polarization direction. Additional calculations in the SI show that, to preserve this supervibronic transition mechanism, only the electronic dipole orientations need to exhibit anisotropy. The magnitude of this energy transfer scales proportionally to $\avg{\cos\theta}^2$, where $\theta$ represents the  angle between the dipole vector of an electronic TLS and the cavity polarization direction.} 
    
    \red{This supervibronic transition mechanism might advance vibronic transition processes that are typically local and short-ranged, such as proton-coupled electron transfer \cite{Hammes-Schiffer2015}, to be nonlocal and collective. Our proposal can be tested by performing UV-vis-pump IR-probe ultrafast experiments with a planar Fabry--P\'erot cavity composed of three layers: one middle layer made of inert materials, one side layer with the aligned electronic TLSs \cite{empicka-Mirek2022,Stemo2022}, as well as the opposite layer utilized to form VSC.  Observing remote vibrational excitations of the vibrational layer after UV-vis pumping of the electronic subsystem would be a strong endorsement of this proposed mechanism. This observation provides advantageous experimental evidence for the dynamical Casimir effect without directly detecting the emission of the Casimir photons, which have a much shorter lifetime than the vibrational excitations. Of course, a negative experimental observation may suggest the limitation of employing the Pauli--Fierz Hamiltonian for describing electronic nonadiabatic dynamics under VSC conditions.}    
    

    \paragraph{Acknowledgments.} This work is supported by  start-up funds from the University of Delaware Department of Physics and Astronomy. 
    The author thanks Abraham Nitzan,   Joseph Subotnik, Adam Dunkelberger, Blake Simpkins, and Jeffrey Owrutsky for useful discussions.


\begin{thebibliography}{66}%
    	\makeatletter
    	\providecommand \@ifxundefined [1]{%
    		\@ifx{#1\undefined}
    	}%
    	\providecommand \@ifnum [1]{%
    		\ifnum #1\expandafter \@firstoftwo
    		\else \expandafter \@secondoftwo
    		\fi
    	}%
    	\providecommand \@ifx [1]{%
    		\ifx #1\expandafter \@firstoftwo
    		\else \expandafter \@secondoftwo
    		\fi
    	}%
    	\providecommand \natexlab [1]{#1}%
    	\providecommand \enquote  [1]{``#1''}%
    	\providecommand \bibnamefont  [1]{#1}%
    	\providecommand \bibfnamefont [1]{#1}%
    	\providecommand \citenamefont [1]{#1}%
    	\providecommand \href@noop [0]{\@secondoftwo}%
    	\providecommand \href [0]{\begingroup \@sanitize@url \@href}%
    	\providecommand \@href[1]{\@@startlink{#1}\@@href}%
    	\providecommand \@@href[1]{\endgroup#1\@@endlink}%
    	\providecommand \@sanitize@url [0]{\catcode `\\12\catcode `\$12\catcode
    		`\&12\catcode `\#12\catcode `\^12\catcode `\_12\catcode `\%12\relax}%
    	\providecommand \@@startlink[1]{}%
    	\providecommand \@@endlink[0]{}%
    	\providecommand \url  [0]{\begingroup\@sanitize@url \@url }%
    	\providecommand \@url [1]{\endgroup\@href {#1}{\urlprefix }}%
    	\providecommand \urlprefix  [0]{URL }%
    	\providecommand \Eprint [0]{\href }%
    	\providecommand \doibase [0]{https://doi.org/}%
    	\providecommand \selectlanguage [0]{\@gobble}%
    	\providecommand \bibinfo  [0]{\@secondoftwo}%
    	\providecommand \bibfield  [0]{\@secondoftwo}%
    	\providecommand \translation [1]{[#1]}%
    	\providecommand \BibitemOpen [0]{}%
    	\providecommand \bibitemStop [0]{}%
    	\providecommand \bibitemNoStop [0]{.\EOS\space}%
    	\providecommand \EOS [0]{\spacefactor3000\relax}%
    	\providecommand \BibitemShut  [1]{\csname bibitem#1\endcsname}%
    	\let\auto@bib@innerbib\@empty
    	\bibitem [{\citenamefont {Dicke}(1954)}]{Dicke1954}%
    	\BibitemOpen
    	\bibfield  {author} {\bibinfo {author} {\bibfnamefont {R.~H.}\ \bibnamefont
    			{Dicke}},\ }\bibfield  {title} {\bibinfo {title} {{Coherence in Spontaneous
    				Radiation Processes}},\ }\href {https://doi.org/10.1103/PhysRev.93.99}
    	{\bibfield  {journal} {\bibinfo  {journal} {Phys. Rev.}\ }\textbf {\bibinfo
    			{volume} {93}},\ \bibinfo {pages} {99} (\bibinfo {year} {1954})}\BibitemShut
    	{NoStop}%
    	\bibitem [{\citenamefont {Gross}\ and\ \citenamefont
    		{Haroche}(1982)}]{Gross1982}%
    	\BibitemOpen
    	\bibfield  {author} {\bibinfo {author} {\bibfnamefont {M.}~\bibnamefont
    			{Gross}}\ and\ \bibinfo {author} {\bibfnamefont {S.}~\bibnamefont
    			{Haroche}},\ }\bibfield  {title} {\bibinfo {title} {{Superradiance: An Essay
    				on the Theory of Collective Spontaneous Emission}},\ }\href
    	{https://doi.org/10.1016/0370-1573(82)90102-8} {\bibfield  {journal}
    		{\bibinfo  {journal} {Phys. Rep.}\ }\textbf {\bibinfo {volume} {93}},\
    		\bibinfo {pages} {301} (\bibinfo {year} {1982})}\BibitemShut {NoStop}%
    	\bibitem [{\citenamefont {Breuer}\ and\ \citenamefont
    		{Petruccione}(2007)}]{Breuer2007}%
    	\BibitemOpen
    	\bibfield  {author} {\bibinfo {author} {\bibfnamefont {H.-P.}\ \bibnamefont
    			{Breuer}}\ and\ \bibinfo {author} {\bibfnamefont {F.}~\bibnamefont
    			{Petruccione}},\ }\href
    	{https://doi.org/10.1093/acprof:oso/9780199213900.001.0001} {\emph {\bibinfo
    			{title} {{The Theory of Open Quantum Systems}}}}\ (\bibinfo  {publisher}
    	{Oxford University Press},\ \bibinfo {address} {New York},\ \bibinfo {year}
    	{2007})\BibitemShut {NoStop}%
    	\bibitem [{\citenamefont {Ribeiro}\ \emph {et~al.}(2018)\citenamefont
    		{Ribeiro}, \citenamefont {Mart{\'{i}}nez-Mart{\'{i}}nez}, \citenamefont {Du},
    		\citenamefont {Campos-Gonzalez-Angulo},\ and\ \citenamefont
    		{Yuen-Zhou}}]{Ribeiro2018}%
    	\BibitemOpen
    	\bibfield  {author} {\bibinfo {author} {\bibfnamefont {R.~F.}\ \bibnamefont
    			{Ribeiro}}, \bibinfo {author} {\bibfnamefont {L.~A.}\ \bibnamefont
    			{Mart{\'{i}}nez-Mart{\'{i}}nez}}, \bibinfo {author} {\bibfnamefont
    			{M.}~\bibnamefont {Du}}, \bibinfo {author} {\bibfnamefont {J.}~\bibnamefont
    			{Campos-Gonzalez-Angulo}},\ and\ \bibinfo {author} {\bibfnamefont
    			{J.}~\bibnamefont {Yuen-Zhou}},\ }\bibfield  {title} {\bibinfo {title}
    		{{Polariton Chemistry: Controlling Molecular Dynamics with Optical
    				Cavities}},\ }\href {https://doi.org/10.1039/C8SC01043A} {\bibfield
    		{journal} {\bibinfo  {journal} {Chem. Sci.}\ }\textbf {\bibinfo {volume}
    			{9}},\ \bibinfo {pages} {6325} (\bibinfo {year} {2018})}\BibitemShut
    	{NoStop}%
    	\bibitem [{\citenamefont {{Frisk Kockum}}\ \emph {et~al.}(2019)\citenamefont
    		{{Frisk Kockum}}, \citenamefont {Miranowicz}, \citenamefont {{De Liberato}},
    		\citenamefont {Savasta},\ and\ \citenamefont {Nori}}]{FriskKockum2019}%
    	\BibitemOpen
    	\bibfield  {author} {\bibinfo {author} {\bibfnamefont {A.}~\bibnamefont
    			{{Frisk Kockum}}}, \bibinfo {author} {\bibfnamefont {A.}~\bibnamefont
    			{Miranowicz}}, \bibinfo {author} {\bibfnamefont {S.}~\bibnamefont {{De
    					Liberato}}}, \bibinfo {author} {\bibfnamefont {S.}~\bibnamefont {Savasta}},\
    		and\ \bibinfo {author} {\bibfnamefont {F.}~\bibnamefont {Nori}},\ }\bibfield
    	{title} {\bibinfo {title} {{Ultrastrong Coupling between Light and Matter}},\
    	}\href {https://doi.org/10.1038/s42254-018-0006-2} {\bibfield  {journal}
    		{\bibinfo  {journal} {Nat. Rev. Phys.}\ }\textbf {\bibinfo {volume} {1}},\
    		\bibinfo {pages} {19} (\bibinfo {year} {2019})}\BibitemShut {NoStop}%
    	\bibitem [{\citenamefont {Herrera}\ and\ \citenamefont
    		{Owrutsky}(2020)}]{Herrera2019}%
    	\BibitemOpen
    	\bibfield  {author} {\bibinfo {author} {\bibfnamefont {F.}~\bibnamefont
    			{Herrera}}\ and\ \bibinfo {author} {\bibfnamefont {J.}~\bibnamefont
    			{Owrutsky}},\ }\bibfield  {title} {\bibinfo {title} {{Molecular Polaritons
    				for Controlling Chemistry with Quantum Optics}},\ }\href
    	{https://doi.org/10.1063/1.5136320} {\bibfield  {journal} {\bibinfo
    			{journal} {J. Chem. Phys.}\ }\textbf {\bibinfo {volume} {152}},\ \bibinfo
    		{pages} {100902} (\bibinfo {year} {2020})}\BibitemShut {NoStop}%
    	\bibitem [{\citenamefont {Nagarajan}\ \emph {et~al.}(2021)\citenamefont
    		{Nagarajan}, \citenamefont {Thomas},\ and\ \citenamefont
    		{Ebbesen}}]{Nagarajan2021}%
    	\BibitemOpen
    	\bibfield  {author} {\bibinfo {author} {\bibfnamefont {K.}~\bibnamefont
    			{Nagarajan}}, \bibinfo {author} {\bibfnamefont {A.}~\bibnamefont {Thomas}},\
    		and\ \bibinfo {author} {\bibfnamefont {T.~W.}\ \bibnamefont {Ebbesen}},\
    	}\bibfield  {title} {\bibinfo {title} {{Chemistry under Vibrational Strong
    				Coupling}},\ }\href {https://doi.org/10.1021/jacs.1c07420} {\bibfield
    		{journal} {\bibinfo  {journal} {J. Am. Chem. Soc.}\ }\textbf {\bibinfo
    			{volume} {143}},\ \bibinfo {pages} {16877} (\bibinfo {year}
    		{2021})}\BibitemShut {NoStop}%
    	\bibitem [{\citenamefont {Xiang}\ and\ \citenamefont
    		{Xiong}(2021)}]{Xiang2021JCP}%
    	\BibitemOpen
    	\bibfield  {author} {\bibinfo {author} {\bibfnamefont {B.}~\bibnamefont
    			{Xiang}}\ and\ \bibinfo {author} {\bibfnamefont {W.}~\bibnamefont {Xiong}},\
    	}\bibfield  {title} {\bibinfo {title} {{Molecular Vibrational Polariton: Its
    				Dynamics and Potentials in Novel Chemistry and Quantum Technology}},\ }\href
    	{https://doi.org/10.1063/5.0054896} {\bibfield  {journal} {\bibinfo
    			{journal} {J. Chem. Phys.}\ }\textbf {\bibinfo {volume} {155}},\ \bibinfo
    		{pages} {050901} (\bibinfo {year} {2021})}\BibitemShut {NoStop}%
    	\bibitem [{\citenamefont {Fregoni}\ \emph {et~al.}(2022)\citenamefont
    		{Fregoni}, \citenamefont {Garcia-Vidal},\ and\ \citenamefont
    		{Feist}}]{Fregoni2022}%
    	\BibitemOpen
    	\bibfield  {author} {\bibinfo {author} {\bibfnamefont {J.}~\bibnamefont
    			{Fregoni}}, \bibinfo {author} {\bibfnamefont {F.~J.}\ \bibnamefont
    			{Garcia-Vidal}},\ and\ \bibinfo {author} {\bibfnamefont {J.}~\bibnamefont
    			{Feist}},\ }\bibfield  {title} {\bibinfo {title} {{Theoretical Challenges in
    				Polaritonic Chemistry}},\ }\href
    	{https://doi.org/10.1021/acsphotonics.1c01749} {\bibfield  {journal}
    		{\bibinfo  {journal} {ACS Photonics}\ }\textbf {\bibinfo {volume} {9}},\
    		\bibinfo {pages} {1096} (\bibinfo {year} {2022})}\BibitemShut {NoStop}%
    	\bibitem [{\citenamefont {Simpkins}\ \emph {et~al.}(2023)\citenamefont
    		{Simpkins}, \citenamefont {Dunkelberger},\ and\ \citenamefont
    		{Vurgaftman}}]{Simpkins2023}%
    	\BibitemOpen
    	\bibfield  {author} {\bibinfo {author} {\bibfnamefont {B.~S.}\ \bibnamefont
    			{Simpkins}}, \bibinfo {author} {\bibfnamefont {A.~D.}\ \bibnamefont
    			{Dunkelberger}},\ and\ \bibinfo {author} {\bibfnamefont {I.}~\bibnamefont
    			{Vurgaftman}},\ }\bibfield  {title} {\bibinfo {title} {{Control, Modulation,
    				and Analytical Descriptions of Vibrational Strong Coupling}},\ }\href
    	{https://doi.org/10.1021/acs.chemrev.2c00774} {\bibfield  {journal} {\bibinfo
    			{journal} {Chem. Rev.}\ }\textbf {\bibinfo {volume} {123}},\ \bibinfo
    		{pages} {5020} (\bibinfo {year} {2023})}\BibitemShut {NoStop}%
    	\bibitem [{\citenamefont {Mandal}\ \emph {et~al.}(2023)\citenamefont {Mandal},
    		\citenamefont {Taylor}, \citenamefont {Weight}, \citenamefont {Koessler},
    		\citenamefont {Li},\ and\ \citenamefont {Huo}}]{Mandal2023ChemRev}%
    	\BibitemOpen
    	\bibfield  {author} {\bibinfo {author} {\bibfnamefont {A.}~\bibnamefont
    			{Mandal}}, \bibinfo {author} {\bibfnamefont {M.~A.}\ \bibnamefont {Taylor}},
    		\bibinfo {author} {\bibfnamefont {B.~M.}\ \bibnamefont {Weight}}, \bibinfo
    		{author} {\bibfnamefont {E.~R.}\ \bibnamefont {Koessler}}, \bibinfo {author}
    		{\bibfnamefont {X.}~\bibnamefont {Li}},\ and\ \bibinfo {author}
    		{\bibfnamefont {P.}~\bibnamefont {Huo}},\ }\bibfield  {title} {\bibinfo
    		{title} {{Theoretical Advances in Polariton Chemistry and Molecular Cavity
    				Quantum Electrodynamics}},\ }\href
    	{https://doi.org/10.1021/acs.chemrev.2c00855} {\bibfield  {journal} {\bibinfo
    			{journal} {Chem. Rev.}\ }\textbf {\bibinfo {volume} {123}},\ \bibinfo
    		{pages} {9786} (\bibinfo {year} {2023})}\BibitemShut {NoStop}%
    	\bibitem [{\citenamefont {Shalabney}\ \emph {et~al.}(2015)\citenamefont
    		{Shalabney}, \citenamefont {George}, \citenamefont {Hutchison}, \citenamefont
    		{Pupillo}, \citenamefont {Genet},\ and\ \citenamefont
    		{Ebbesen}}]{Shalabney2015}%
    	\BibitemOpen
    	\bibfield  {author} {\bibinfo {author} {\bibfnamefont {A.}~\bibnamefont
    			{Shalabney}}, \bibinfo {author} {\bibfnamefont {J.}~\bibnamefont {George}},
    		\bibinfo {author} {\bibfnamefont {J.}~\bibnamefont {Hutchison}}, \bibinfo
    		{author} {\bibfnamefont {G.}~\bibnamefont {Pupillo}}, \bibinfo {author}
    		{\bibfnamefont {C.}~\bibnamefont {Genet}},\ and\ \bibinfo {author}
    		{\bibfnamefont {T.~W.}\ \bibnamefont {Ebbesen}},\ }\bibfield  {title}
    	{\bibinfo {title} {{Coherent Coupling of Molecular Resonators with a
    				Microcavity Mode}},\ }\href {https://doi.org/10.1038/ncomms6981} {\bibfield
    		{journal} {\bibinfo  {journal} {Nat. Commun.}\ }\textbf {\bibinfo {volume}
    			{6}},\ \bibinfo {pages} {5981} (\bibinfo {year} {2015})}\BibitemShut
    	{NoStop}%
    	\bibitem [{\citenamefont {Long}\ and\ \citenamefont
    		{Simpkins}(2015)}]{Long2015}%
    	\BibitemOpen
    	\bibfield  {author} {\bibinfo {author} {\bibfnamefont {J.~P.}\ \bibnamefont
    			{Long}}\ and\ \bibinfo {author} {\bibfnamefont {B.~S.}\ \bibnamefont
    			{Simpkins}},\ }\bibfield  {title} {\bibinfo {title} {{Coherent Coupling
    				between a Molecular Vibration and Fabry–Perot Optical Cavity to Give
    				Hybridized States in the Strong Coupling Limit}},\ }\href
    	{https://doi.org/10.1021/ph5003347} {\bibfield  {journal} {\bibinfo
    			{journal} {ACS Photonics}\ }\textbf {\bibinfo {volume} {2}},\ \bibinfo
    		{pages} {130} (\bibinfo {year} {2015})}\BibitemShut {NoStop}%
    	\bibitem [{\citenamefont {Thomas}\ \emph {et~al.}(2016)\citenamefont {Thomas},
    		\citenamefont {George}, \citenamefont {Shalabney}, \citenamefont {Dryzhakov},
    		\citenamefont {Varma}, \citenamefont {Moran}, \citenamefont {Chervy},
    		\citenamefont {Zhong}, \citenamefont {Devaux}, \citenamefont {Genet},
    		\citenamefont {Hutchison},\ and\ \citenamefont {Ebbesen}}]{Thomas2016}%
    	\BibitemOpen
    	\bibfield  {author} {\bibinfo {author} {\bibfnamefont {A.}~\bibnamefont
    			{Thomas}}, \bibinfo {author} {\bibfnamefont {J.}~\bibnamefont {George}},
    		\bibinfo {author} {\bibfnamefont {A.}~\bibnamefont {Shalabney}}, \bibinfo
    		{author} {\bibfnamefont {M.}~\bibnamefont {Dryzhakov}}, \bibinfo {author}
    		{\bibfnamefont {S.~J.}\ \bibnamefont {Varma}}, \bibinfo {author}
    		{\bibfnamefont {J.}~\bibnamefont {Moran}}, \bibinfo {author} {\bibfnamefont
    			{T.}~\bibnamefont {Chervy}}, \bibinfo {author} {\bibfnamefont
    			{X.}~\bibnamefont {Zhong}}, \bibinfo {author} {\bibfnamefont
    			{E.}~\bibnamefont {Devaux}}, \bibinfo {author} {\bibfnamefont
    			{C.}~\bibnamefont {Genet}}, \bibinfo {author} {\bibfnamefont {J.~A.}\
    			\bibnamefont {Hutchison}},\ and\ \bibinfo {author} {\bibfnamefont {T.~W.}\
    			\bibnamefont {Ebbesen}},\ }\bibfield  {title} {\bibinfo {title}
    		{{Ground-State Chemical Reactivity under Vibrational Coupling to the Vacuum
    				Electromagnetic Field}},\ }\href {https://doi.org/10.1002/anie.201605504}
    	{\bibfield  {journal} {\bibinfo  {journal} {Angew. Chemie Int. Ed.}\ }\textbf
    		{\bibinfo {volume} {55}},\ \bibinfo {pages} {11462} (\bibinfo {year}
    		{2016})}\BibitemShut {NoStop}%
    	\bibitem [{\citenamefont {Thomas}\ \emph {et~al.}(2019)\citenamefont {Thomas},
    		\citenamefont {Lethuillier-Karl}, \citenamefont {Nagarajan}, \citenamefont
    		{Vergauwe}, \citenamefont {George}, \citenamefont {Chervy}, \citenamefont
    		{Shalabney}, \citenamefont {Devaux}, \citenamefont {Genet}, \citenamefont
    		{Moran},\ and\ \citenamefont {Ebbesen}}]{Thomas2019_science}%
    	\BibitemOpen
    	\bibfield  {author} {\bibinfo {author} {\bibfnamefont {A.}~\bibnamefont
    			{Thomas}}, \bibinfo {author} {\bibfnamefont {L.}~\bibnamefont
    			{Lethuillier-Karl}}, \bibinfo {author} {\bibfnamefont {K.}~\bibnamefont
    			{Nagarajan}}, \bibinfo {author} {\bibfnamefont {R.~M.~A.}\ \bibnamefont
    			{Vergauwe}}, \bibinfo {author} {\bibfnamefont {J.}~\bibnamefont {George}},
    		\bibinfo {author} {\bibfnamefont {T.}~\bibnamefont {Chervy}}, \bibinfo
    		{author} {\bibfnamefont {A.}~\bibnamefont {Shalabney}}, \bibinfo {author}
    		{\bibfnamefont {E.}~\bibnamefont {Devaux}}, \bibinfo {author} {\bibfnamefont
    			{C.}~\bibnamefont {Genet}}, \bibinfo {author} {\bibfnamefont
    			{J.}~\bibnamefont {Moran}},\ and\ \bibinfo {author} {\bibfnamefont {T.~W.}\
    			\bibnamefont {Ebbesen}},\ }\bibfield  {title} {\bibinfo {title} {{Tilting a
    				Ground-State Reactivity Landscape by Vibrational Strong Coupling}},\ }\href
    	{https://doi.org/10.1126/science.aau7742} {\bibfield  {journal} {\bibinfo
    			{journal} {Science}\ }\textbf {\bibinfo {volume} {363}},\ \bibinfo {pages}
    		{615} (\bibinfo {year} {2019})}\BibitemShut {NoStop}%
    	\bibitem [{\citenamefont {Thomas}\ \emph {et~al.}(2020)\citenamefont {Thomas},
    		\citenamefont {Jayachandran}, \citenamefont {Lethuillier-Karl}, \citenamefont
    		{Vergauwe}, \citenamefont {Nagarajan}, \citenamefont {Devaux}, \citenamefont
    		{Genet}, \citenamefont {Moran},\ and\ \citenamefont {Ebbesen}}]{Thomas2020}%
    	\BibitemOpen
    	\bibfield  {author} {\bibinfo {author} {\bibfnamefont {A.}~\bibnamefont
    			{Thomas}}, \bibinfo {author} {\bibfnamefont {A.}~\bibnamefont
    			{Jayachandran}}, \bibinfo {author} {\bibfnamefont {L.}~\bibnamefont
    			{Lethuillier-Karl}}, \bibinfo {author} {\bibfnamefont {R.~M.}\ \bibnamefont
    			{Vergauwe}}, \bibinfo {author} {\bibfnamefont {K.}~\bibnamefont {Nagarajan}},
    		\bibinfo {author} {\bibfnamefont {E.}~\bibnamefont {Devaux}}, \bibinfo
    		{author} {\bibfnamefont {C.}~\bibnamefont {Genet}}, \bibinfo {author}
    		{\bibfnamefont {J.}~\bibnamefont {Moran}},\ and\ \bibinfo {author}
    		{\bibfnamefont {T.~W.}\ \bibnamefont {Ebbesen}},\ }\bibfield  {title}
    	{\bibinfo {title} {{Ground State Chemistry under Vibrational Strong Coupling:
    				Dependence of Thermodynamic Parameters on the Rabi Splitting Energy}},\
    	}\href {https://doi.org/10.1515/nanoph-2019-0340} {\bibfield  {journal}
    		{\bibinfo  {journal} {Nanophoton.}\ }\textbf {\bibinfo {volume} {9}},\
    		\bibinfo {pages} {249} (\bibinfo {year} {2020})}\BibitemShut {NoStop}%
    	\bibitem [{\citenamefont {Ahn}\ \emph {et~al.}(2023)\citenamefont {Ahn},
    		\citenamefont {Triana}, \citenamefont {Recabal}, \citenamefont {Herrera},\
    		and\ \citenamefont {Simpkins}}]{Ahn2023Science}%
    	\BibitemOpen
    	\bibfield  {author} {\bibinfo {author} {\bibfnamefont {W.}~\bibnamefont
    			{Ahn}}, \bibinfo {author} {\bibfnamefont {J.~F.}\ \bibnamefont {Triana}},
    		\bibinfo {author} {\bibfnamefont {F.}~\bibnamefont {Recabal}}, \bibinfo
    		{author} {\bibfnamefont {F.}~\bibnamefont {Herrera}},\ and\ \bibinfo {author}
    		{\bibfnamefont {B.~S.}\ \bibnamefont {Simpkins}},\ }\bibfield  {title}
    	{\bibinfo {title} {{Modification of ground-state chemical reactivity via
    				light–matter coherence in infrared cavities}},\ }\href
    	{https://doi.org/10.1126/science.ade7147} {\bibfield  {journal} {\bibinfo
    			{journal} {Science}\ }\textbf {\bibinfo {volume} {380}},\ \bibinfo {pages}
    		{1165} (\bibinfo {year} {2023})}\BibitemShut {NoStop}%
    	\bibitem [{\citenamefont {Coles}\ \emph {et~al.}(2014)\citenamefont {Coles},
    		\citenamefont {Somaschi}, \citenamefont {Michetti}, \citenamefont {Clark},
    		\citenamefont {Lagoudakis}, \citenamefont {Savvidis},\ and\ \citenamefont
    		{Lidzey}}]{Coles2014}%
    	\BibitemOpen
    	\bibfield  {author} {\bibinfo {author} {\bibfnamefont {D.~M.}\ \bibnamefont
    			{Coles}}, \bibinfo {author} {\bibfnamefont {N.}~\bibnamefont {Somaschi}},
    		\bibinfo {author} {\bibfnamefont {P.}~\bibnamefont {Michetti}}, \bibinfo
    		{author} {\bibfnamefont {C.}~\bibnamefont {Clark}}, \bibinfo {author}
    		{\bibfnamefont {P.~G.}\ \bibnamefont {Lagoudakis}}, \bibinfo {author}
    		{\bibfnamefont {P.~G.}\ \bibnamefont {Savvidis}},\ and\ \bibinfo {author}
    		{\bibfnamefont {D.~G.}\ \bibnamefont {Lidzey}},\ }\bibfield  {title}
    	{\bibinfo {title} {{Polariton-Mediated Energy Transfer between Organic Dyes
    				in A Strongly Coupled Optical Microcavity}},\ }\href
    	{https://doi.org/10.1038/nmat3950} {\bibfield  {journal} {\bibinfo  {journal}
    			{Nat. Mater.}\ }\textbf {\bibinfo {volume} {13}},\ \bibinfo {pages} {712}
    		(\bibinfo {year} {2014})}\BibitemShut {NoStop}%
    	\bibitem [{\citenamefont {Xiang}\ \emph {et~al.}(2020)\citenamefont {Xiang},
    		\citenamefont {Ribeiro}, \citenamefont {Du}, \citenamefont {Chen},
    		\citenamefont {Yang}, \citenamefont {Wang}, \citenamefont {Yuen-Zhou},\ and\
    		\citenamefont {Xiong}}]{Xiang2020Science}%
    	\BibitemOpen
    	\bibfield  {author} {\bibinfo {author} {\bibfnamefont {B.}~\bibnamefont
    			{Xiang}}, \bibinfo {author} {\bibfnamefont {R.~F.}\ \bibnamefont {Ribeiro}},
    		\bibinfo {author} {\bibfnamefont {M.}~\bibnamefont {Du}}, \bibinfo {author}
    		{\bibfnamefont {L.}~\bibnamefont {Chen}}, \bibinfo {author} {\bibfnamefont
    			{Z.}~\bibnamefont {Yang}}, \bibinfo {author} {\bibfnamefont {J.}~\bibnamefont
    			{Wang}}, \bibinfo {author} {\bibfnamefont {J.}~\bibnamefont {Yuen-Zhou}},\
    		and\ \bibinfo {author} {\bibfnamefont {W.}~\bibnamefont {Xiong}},\ }\bibfield
    	{title} {\bibinfo {title} {{Intermolecular Vibrational Energy Transfer
    				Enabled by Microcavity Strong Light--Matter Coupling}},\ }\href
    	{https://doi.org/10.1126/science.aba3544} {\bibfield  {journal} {\bibinfo
    			{journal} {Science}\ }\textbf {\bibinfo {volume} {368}},\ \bibinfo {pages}
    		{665} (\bibinfo {year} {2020})}\BibitemShut {NoStop}%
    	\bibitem [{\citenamefont {Grafton}\ \emph {et~al.}(2021)\citenamefont
    		{Grafton}, \citenamefont {Dunkelberger}, \citenamefont {Simpkins},
    		\citenamefont {Triana}, \citenamefont {Hern{\'{a}}ndez}, \citenamefont
    		{Herrera},\ and\ \citenamefont {Owrutsky}}]{Grafton2020}%
    	\BibitemOpen
    	\bibfield  {author} {\bibinfo {author} {\bibfnamefont {A.~B.}\ \bibnamefont
    			{Grafton}}, \bibinfo {author} {\bibfnamefont {A.~D.}\ \bibnamefont
    			{Dunkelberger}}, \bibinfo {author} {\bibfnamefont {B.~S.}\ \bibnamefont
    			{Simpkins}}, \bibinfo {author} {\bibfnamefont {J.~F.}\ \bibnamefont
    			{Triana}}, \bibinfo {author} {\bibfnamefont {F.~J.}\ \bibnamefont
    			{Hern{\'{a}}ndez}}, \bibinfo {author} {\bibfnamefont {F.}~\bibnamefont
    			{Herrera}},\ and\ \bibinfo {author} {\bibfnamefont {J.~C.}\ \bibnamefont
    			{Owrutsky}},\ }\bibfield  {title} {\bibinfo {title} {{Excited-State
    				Vibration-Polariton Transitions and Dynamics in Nitroprusside}},\ }\href
    	{https://doi.org/10.1038/s41467-020-20535-z} {\bibfield  {journal} {\bibinfo
    			{journal} {Nat. Commun.}\ }\textbf {\bibinfo {volume} {12}},\ \bibinfo
    		{pages} {214} (\bibinfo {year} {2021})}\BibitemShut {NoStop}%
    	\bibitem [{\citenamefont {Polak}\ \emph {et~al.}(2020)\citenamefont {Polak},
    		\citenamefont {Jayaprakash}, \citenamefont {Lyons}, \citenamefont
    		{Mart{\'{i}}nez-Mart{\'{i}}nez}, \citenamefont {Leventis}, \citenamefont
    		{Fallon}, \citenamefont {Coulthard}, \citenamefont {Bossanyi}, \citenamefont
    		{Georgiou}, \citenamefont {{Petty, II}}, \citenamefont {Anthony},
    		\citenamefont {Bronstein}, \citenamefont {Yuen-Zhou}, \citenamefont
    		{Tartakovskii}, \citenamefont {Clark},\ and\ \citenamefont
    		{Musser}}]{Polak2020}%
    	\BibitemOpen
    	\bibfield  {author} {\bibinfo {author} {\bibfnamefont {D.}~\bibnamefont
    			{Polak}}, \bibinfo {author} {\bibfnamefont {R.}~\bibnamefont {Jayaprakash}},
    		\bibinfo {author} {\bibfnamefont {T.~P.}\ \bibnamefont {Lyons}}, \bibinfo
    		{author} {\bibfnamefont {L.~{\'{A}}.}\ \bibnamefont
    			{Mart{\'{i}}nez-Mart{\'{i}}nez}}, \bibinfo {author} {\bibfnamefont
    			{A.}~\bibnamefont {Leventis}}, \bibinfo {author} {\bibfnamefont {K.~J.}\
    			\bibnamefont {Fallon}}, \bibinfo {author} {\bibfnamefont {H.}~\bibnamefont
    			{Coulthard}}, \bibinfo {author} {\bibfnamefont {D.~G.}\ \bibnamefont
    			{Bossanyi}}, \bibinfo {author} {\bibfnamefont {K.}~\bibnamefont {Georgiou}},
    		\bibinfo {author} {\bibfnamefont {A.~J.}\ \bibnamefont {{Petty, II}}},
    		\bibinfo {author} {\bibfnamefont {J.}~\bibnamefont {Anthony}}, \bibinfo
    		{author} {\bibfnamefont {H.}~\bibnamefont {Bronstein}}, \bibinfo {author}
    		{\bibfnamefont {J.}~\bibnamefont {Yuen-Zhou}}, \bibinfo {author}
    		{\bibfnamefont {A.~I.}\ \bibnamefont {Tartakovskii}}, \bibinfo {author}
    		{\bibfnamefont {J.}~\bibnamefont {Clark}},\ and\ \bibinfo {author}
    		{\bibfnamefont {A.~J.}\ \bibnamefont {Musser}},\ }\bibfield  {title}
    	{\bibinfo {title} {{Manipulating Molecules with Strong Coupling: Harvesting
    				Triplet Excitons in Organic Exciton Microcavities}},\ }\href
    	{https://doi.org/10.1039/C9SC04950A} {\bibfield  {journal} {\bibinfo
    			{journal} {Chem. Sci.}\ }\textbf {\bibinfo {volume} {11}},\ \bibinfo {pages}
    		{343} (\bibinfo {year} {2020})}\BibitemShut {NoStop}%
    	\bibitem [{\citenamefont {Georgiou}\ \emph {et~al.}(2021)\citenamefont
    		{Georgiou}, \citenamefont {Jayaprakash}, \citenamefont {Othonos},\ and\
    		\citenamefont {Lidzey}}]{Georgiou2021}%
    	\BibitemOpen
    	\bibfield  {author} {\bibinfo {author} {\bibfnamefont {K.}~\bibnamefont
    			{Georgiou}}, \bibinfo {author} {\bibfnamefont {R.}~\bibnamefont
    			{Jayaprakash}}, \bibinfo {author} {\bibfnamefont {A.}~\bibnamefont
    			{Othonos}},\ and\ \bibinfo {author} {\bibfnamefont {D.~G.}\ \bibnamefont
    			{Lidzey}},\ }\bibfield  {title} {\bibinfo {title} {{Ultralong-Range
    				Polariton-Assisted Energy Transfer in Organic Microcavities}},\ }\href
    	{https://doi.org/10.1002/ANIE.202105442} {\bibfield  {journal} {\bibinfo
    			{journal} {Angew. Chemie Int. Ed.}\ }\textbf {\bibinfo {volume} {60}},\
    		\bibinfo {pages} {16661} (\bibinfo {year} {2021})}\BibitemShut {NoStop}%
    	\bibitem [{\citenamefont {Chen}\ \emph {et~al.}(2022)\citenamefont {Chen},
    		\citenamefont {Du}, \citenamefont {Yang}, \citenamefont {Yuen-Zhou},\ and\
    		\citenamefont {Xiong}}]{Chen2022}%
    	\BibitemOpen
    	\bibfield  {author} {\bibinfo {author} {\bibfnamefont {T.-T.}\ \bibnamefont
    			{Chen}}, \bibinfo {author} {\bibfnamefont {M.}~\bibnamefont {Du}}, \bibinfo
    		{author} {\bibfnamefont {Z.}~\bibnamefont {Yang}}, \bibinfo {author}
    		{\bibfnamefont {J.}~\bibnamefont {Yuen-Zhou}},\ and\ \bibinfo {author}
    		{\bibfnamefont {W.}~\bibnamefont {Xiong}},\ }\bibfield  {title} {\bibinfo
    		{title} {{Cavity-enabled Enhancement of Ultrafast Intramolecular Vibrational
    				Redistribution over Pseudorotation}},\ }\href
    	{https://doi.org/10.1126/science.add0276} {\bibfield  {journal} {\bibinfo
    			{journal} {Science}\ }\textbf {\bibinfo {volume} {378}},\ \bibinfo {pages}
    		{790} (\bibinfo {year} {2022})}\BibitemShut {NoStop}%
    	\bibitem [{\citenamefont {Pino}\ \emph {et~al.}(2015)\citenamefont {Pino},
    		\citenamefont {Feist},\ and\ \citenamefont {Garcia-Vidal}}]{Pino2015}%
    	\BibitemOpen
    	\bibfield  {author} {\bibinfo {author} {\bibfnamefont {J.~D.}\ \bibnamefont
    			{Pino}}, \bibinfo {author} {\bibfnamefont {J.}~\bibnamefont {Feist}},\ and\
    		\bibinfo {author} {\bibfnamefont {F.~J.}\ \bibnamefont {Garcia-Vidal}},\
    	}\bibfield  {title} {\bibinfo {title} {{Quantum Theory of Collective Strong
    				Coupling of Molecular Vibrations with a Microcavity Mode}},\ }\href
    	{https://doi.org/10.1088/1367-2630/17/5/053040} {\bibfield  {journal}
    		{\bibinfo  {journal} {New J. Phys.}\ }\textbf {\bibinfo {volume} {17}},\
    		\bibinfo {pages} {053040} (\bibinfo {year} {2015})}\BibitemShut {NoStop}%
    	\bibitem [{\citenamefont {Du}\ \emph {et~al.}(2018)\citenamefont {Du},
    		\citenamefont {Mart{\'{i}}nez-Mart{\'{i}}nez}, \citenamefont {Ribeiro},
    		\citenamefont {Hu}, \citenamefont {Menon},\ and\ \citenamefont
    		{Yuen-Zhou}}]{Du2018}%
    	\BibitemOpen
    	\bibfield  {author} {\bibinfo {author} {\bibfnamefont {M.}~\bibnamefont
    			{Du}}, \bibinfo {author} {\bibfnamefont {L.~A.}\ \bibnamefont
    			{Mart{\'{i}}nez-Mart{\'{i}}nez}}, \bibinfo {author} {\bibfnamefont {R.~F.}\
    			\bibnamefont {Ribeiro}}, \bibinfo {author} {\bibfnamefont {Z.}~\bibnamefont
    			{Hu}}, \bibinfo {author} {\bibfnamefont {V.~M.}\ \bibnamefont {Menon}},\ and\
    		\bibinfo {author} {\bibfnamefont {J.}~\bibnamefont {Yuen-Zhou}},\ }\bibfield
    	{title} {\bibinfo {title} {{Theory for Polariton-Assisted Remote Energy
    				Transfer}},\ }\href {https://doi.org/10.1039/C8SC00171E} {\bibfield
    		{journal} {\bibinfo  {journal} {Chem. Sci.}\ }\textbf {\bibinfo {volume}
    			{9}},\ \bibinfo {pages} {6659} (\bibinfo {year} {2018})}\BibitemShut
    	{NoStop}%
    	\bibitem [{\citenamefont {S{\'{a}}ez-Bl{\'{a}}zquez}\ \emph
    		{et~al.}(2018)\citenamefont {S{\'{a}}ez-Bl{\'{a}}zquez}, \citenamefont
    		{Feist}, \citenamefont {Fern{\'{a}}ndez-Dom{\'{i}}nguez},\ and\ \citenamefont
    		{Garc{\'{i}}a-Vidal}}]{Saez-Blazquez2018}%
    	\BibitemOpen
    	\bibfield  {author} {\bibinfo {author} {\bibfnamefont {R.}~\bibnamefont
    			{S{\'{a}}ez-Bl{\'{a}}zquez}}, \bibinfo {author} {\bibfnamefont
    			{J.}~\bibnamefont {Feist}}, \bibinfo {author} {\bibfnamefont {A.~I.}\
    			\bibnamefont {Fern{\'{a}}ndez-Dom{\'{i}}nguez}},\ and\ \bibinfo {author}
    		{\bibfnamefont {F.~J.}\ \bibnamefont {Garc{\'{i}}a-Vidal}},\ }\bibfield
    	{title} {\bibinfo {title} {{Organic Polaritons Enable Local Vibrations to
    				Drive Long-Range Energy Transfer}},\ }\href
    	{https://doi.org/10.1103/PHYSREVB.97.241407/FIGURES/3/MEDIUM} {\bibfield
    		{journal} {\bibinfo  {journal} {Phys. Rev. B}\ }\textbf {\bibinfo {volume}
    			{97}},\ \bibinfo {pages} {241407} (\bibinfo {year} {2018})}\BibitemShut
    	{NoStop}%
    	\bibitem [{\citenamefont {Galego}\ \emph {et~al.}(2019)\citenamefont {Galego},
    		\citenamefont {Climent}, \citenamefont {Garcia-Vidal},\ and\ \citenamefont
    		{Feist}}]{Galego2019}%
    	\BibitemOpen
    	\bibfield  {author} {\bibinfo {author} {\bibfnamefont {J.}~\bibnamefont
    			{Galego}}, \bibinfo {author} {\bibfnamefont {C.}~\bibnamefont {Climent}},
    		\bibinfo {author} {\bibfnamefont {F.~J.}\ \bibnamefont {Garcia-Vidal}},\ and\
    		\bibinfo {author} {\bibfnamefont {J.}~\bibnamefont {Feist}},\ }\bibfield
    	{title} {\bibinfo {title} {{Cavity Casimir-Polder Forces and Their Effects in
    				Ground-State Chemical Reactivity}},\ }\href
    	{https://doi.org/10.1103/PhysRevX.9.021057} {\bibfield  {journal} {\bibinfo
    			{journal} {Phys. Rev. X}\ }\textbf {\bibinfo {volume} {9}},\ \bibinfo {pages}
    		{021057} (\bibinfo {year} {2019})}\BibitemShut {NoStop}%
    	\bibitem [{\citenamefont {Semenov}\ and\ \citenamefont
    		{Nitzan}(2019)}]{Semenov2019}%
    	\BibitemOpen
    	\bibfield  {author} {\bibinfo {author} {\bibfnamefont {A.}~\bibnamefont
    			{Semenov}}\ and\ \bibinfo {author} {\bibfnamefont {A.}~\bibnamefont
    			{Nitzan}},\ }\bibfield  {title} {\bibinfo {title} {{Electron Transfer in
    				Confined Electromagnetic Fields}},\ }\href
    	{https://doi.org/10.1063/1.5095940/16743342} {\bibfield  {journal} {\bibinfo
    			{journal} {J. Chem. Phys.}\ }\textbf {\bibinfo {volume} {150}},\ \bibinfo
    		{pages} {174122} (\bibinfo {year} {2019})}\BibitemShut {NoStop}%
    	\bibitem [{\citenamefont {Climent}\ \emph {et~al.}(2019)\citenamefont
    		{Climent}, \citenamefont {Galego}, \citenamefont {Garcia‐Vidal},\ and\
    		\citenamefont {Feist}}]{Climent2019Plasmonic}%
    	\BibitemOpen
    	\bibfield  {author} {\bibinfo {author} {\bibfnamefont {C.}~\bibnamefont
    			{Climent}}, \bibinfo {author} {\bibfnamefont {J.}~\bibnamefont {Galego}},
    		\bibinfo {author} {\bibfnamefont {F.~J.}\ \bibnamefont {Garcia‐Vidal}},\
    		and\ \bibinfo {author} {\bibfnamefont {J.}~\bibnamefont {Feist}},\ }\bibfield
    	{title} {\bibinfo {title} {{Plasmonic Nanocavities Enable Self‐Induced
    				Electrostatic Catalysis}},\ }\href {https://doi.org/10.1002/anie.201901926}
    	{\bibfield  {journal} {\bibinfo  {journal} {Angew. Chemie Int. Ed.}\ }\textbf
    		{\bibinfo {volume} {58}},\ \bibinfo {pages} {8698} (\bibinfo {year}
    		{2019})}\BibitemShut {NoStop}%
    	\bibitem [{\citenamefont {Lacombe}\ \emph {et~al.}(2019)\citenamefont
    		{Lacombe}, \citenamefont {Hoffmann},\ and\ \citenamefont
    		{Maitra}}]{Lacombe2019}%
    	\BibitemOpen
    	\bibfield  {author} {\bibinfo {author} {\bibfnamefont {L.}~\bibnamefont
    			{Lacombe}}, \bibinfo {author} {\bibfnamefont {N.~M.}\ \bibnamefont
    			{Hoffmann}},\ and\ \bibinfo {author} {\bibfnamefont {N.~T.}\ \bibnamefont
    			{Maitra}},\ }\bibfield  {title} {\bibinfo {title} {{Exact Potential Energy
    				Surface for Molecules in Cavities}},\ }\href
    	{https://doi.org/10.1103/PhysRevLett.123.083201} {\bibfield  {journal}
    		{\bibinfo  {journal} {Phys. Rev. Lett.}\ }\textbf {\bibinfo {volume} {123}},\
    		\bibinfo {pages} {083201} (\bibinfo {year} {2019})},\ \Eprint
    	{https://arxiv.org/abs/1906.02651} {arXiv:1906.02651} \BibitemShut {NoStop}%
    	\bibitem [{\citenamefont {Hoffmann}\ \emph {et~al.}(2020)\citenamefont
    		{Hoffmann}, \citenamefont {Lacombe}, \citenamefont {Rubio},\ and\
    		\citenamefont {Maitra}}]{Hoffmann2020}%
    	\BibitemOpen
    	\bibfield  {author} {\bibinfo {author} {\bibfnamefont {N.~M.}\ \bibnamefont
    			{Hoffmann}}, \bibinfo {author} {\bibfnamefont {L.}~\bibnamefont {Lacombe}},
    		\bibinfo {author} {\bibfnamefont {A.}~\bibnamefont {Rubio}},\ and\ \bibinfo
    		{author} {\bibfnamefont {N.~T.}\ \bibnamefont {Maitra}},\ }\bibfield  {title}
    	{\bibinfo {title} {{Effect of Many Modes on Self-Polarization and
    				Photochemical Suppression in Cavities}},\ }\href
    	{https://doi.org/10.1063/5.0012723} {\bibfield  {journal} {\bibinfo
    			{journal} {J. Chem. Phys.}\ }\textbf {\bibinfo {volume} {153}},\ \bibinfo
    		{pages} {104103} (\bibinfo {year} {2020})}\BibitemShut {NoStop}%
    	\bibitem [{\citenamefont {Li}\ \emph {et~al.}(2020)\citenamefont {Li},
    		\citenamefont {Nitzan},\ and\ \citenamefont {Subotnik}}]{Li2020Origin}%
    	\BibitemOpen
    	\bibfield  {author} {\bibinfo {author} {\bibfnamefont {T.~E.}\ \bibnamefont
    			{Li}}, \bibinfo {author} {\bibfnamefont {A.}~\bibnamefont {Nitzan}},\ and\
    		\bibinfo {author} {\bibfnamefont {J.~E.}\ \bibnamefont {Subotnik}},\
    	}\bibfield  {title} {\bibinfo {title} {{On the Origin of Ground-State
    				Vacuum-Field Catalysis: Equilibrium Consideration}},\ }\href
    	{https://doi.org/10.1063/5.0006472} {\bibfield  {journal} {\bibinfo
    			{journal} {J. Chem. Phys.}\ }\textbf {\bibinfo {volume} {152}},\ \bibinfo
    		{pages} {234107} (\bibinfo {year} {2020})}\BibitemShut {NoStop}%
    	\bibitem [{\citenamefont {P{\'{e}}rez-S{\'{a}}nchez}\ and\ \citenamefont
    		{Yuen-Zhou}(2020)}]{Perez-Sanchez2020}%
    	\BibitemOpen
    	\bibfield  {author} {\bibinfo {author} {\bibfnamefont {J.~B.}\ \bibnamefont
    			{P{\'{e}}rez-S{\'{a}}nchez}}\ and\ \bibinfo {author} {\bibfnamefont
    			{J.}~\bibnamefont {Yuen-Zhou}},\ }\bibfield  {title} {\bibinfo {title}
    		{{Polariton Assisted Down-Conversion of Photons via Nonadiabatic Molecular
    				Dynamics: A Molecular Dynamical Casimir Effect}},\ }\href
    	{https://doi.org/10.1021/acs.jpclett.9b02870} {\bibfield  {journal} {\bibinfo
    			{journal} {J. Phys. Chem. Lett.}\ }\textbf {\bibinfo {volume} {11}},\
    		\bibinfo {pages} {152} (\bibinfo {year} {2020})},\ \Eprint
    	{https://arxiv.org/abs/1909.13024} {arXiv:1909.13024} \BibitemShut {NoStop}%
    	\bibitem [{\citenamefont {Li}\ \emph {et~al.}(2021{\natexlab{a}})\citenamefont
    		{Li}, \citenamefont {Mandal},\ and\ \citenamefont {Huo}}]{LiHuo2021}%
    	\BibitemOpen
    	\bibfield  {author} {\bibinfo {author} {\bibfnamefont {X.}~\bibnamefont
    			{Li}}, \bibinfo {author} {\bibfnamefont {A.}~\bibnamefont {Mandal}},\ and\
    		\bibinfo {author} {\bibfnamefont {P.}~\bibnamefont {Huo}},\ }\bibfield
    	{title} {\bibinfo {title} {{Cavity Frequency-Dependent Theory for Vibrational
    				Polariton Chemistry}},\ }\href {https://doi.org/10.1038/s41467-021-21610-9}
    	{\bibfield  {journal} {\bibinfo  {journal} {Nat. Commun.}\ }\textbf {\bibinfo
    			{volume} {12}},\ \bibinfo {pages} {1315} (\bibinfo {year}
    		{2021}{\natexlab{a}})}\BibitemShut {NoStop}%
    	\bibitem [{\citenamefont {Du}\ and\ \citenamefont {Yuen-Zhou}(2022)}]{Du2022}%
    	\BibitemOpen
    	\bibfield  {author} {\bibinfo {author} {\bibfnamefont {M.}~\bibnamefont
    			{Du}}\ and\ \bibinfo {author} {\bibfnamefont {J.}~\bibnamefont {Yuen-Zhou}},\
    	}\bibfield  {title} {\bibinfo {title} {{Catalysis by Dark States in
    				Vibropolaritonic Chemistry}},\ }\href
    	{https://doi.org/10.1103/PhysRevLett.128.096001} {\bibfield  {journal}
    		{\bibinfo  {journal} {Phys. Rev. Lett.}\ }\textbf {\bibinfo {volume} {128}},\
    		\bibinfo {pages} {096001} (\bibinfo {year} {2022})},\ \Eprint
    	{https://arxiv.org/abs/2104.07214} {arXiv:2104.07214} \BibitemShut {NoStop}%
    	\bibitem [{\citenamefont {Mandal}\ \emph {et~al.}(2020)\citenamefont {Mandal},
    		\citenamefont {{Montillo Vega}},\ and\ \citenamefont
    		{Huo}}]{Mandal2020Polarized}%
    	\BibitemOpen
    	\bibfield  {author} {\bibinfo {author} {\bibfnamefont {A.}~\bibnamefont
    			{Mandal}}, \bibinfo {author} {\bibfnamefont {S.}~\bibnamefont {{Montillo
    					Vega}}},\ and\ \bibinfo {author} {\bibfnamefont {P.}~\bibnamefont {Huo}},\
    	}\bibfield  {title} {\bibinfo {title} {{Polarized Fock States and the
    				Dynamical Casimir Effect in Molecular Cavity Quantum Electrodynamics}},\
    	}\href
    	{https://doi.org/10.1021/ACS.JPCLETT.0C02399/SUPPL_FILE/JZ0C02399_SI_001.PDF}
    	{\bibfield  {journal} {\bibinfo  {journal} {J. Phys. Chem. Lett.}\ }\textbf
    		{\bibinfo {volume} {11}},\ \bibinfo {pages} {9215} (\bibinfo {year}
    		{2020})}\BibitemShut {NoStop}%
    	\bibitem [{\citenamefont {Li}\ \emph {et~al.}(2022)\citenamefont {Li},
    		\citenamefont {Nitzan},\ and\ \citenamefont {Subotnik}}]{Li2021Solute}%
    	\BibitemOpen
    	\bibfield  {author} {\bibinfo {author} {\bibfnamefont {T.~E.}\ \bibnamefont
    			{Li}}, \bibinfo {author} {\bibfnamefont {A.}~\bibnamefont {Nitzan}},\ and\
    		\bibinfo {author} {\bibfnamefont {J.~E.}\ \bibnamefont {Subotnik}},\
    	}\bibfield  {title} {\bibinfo {title} {{Energy-Efficient Pathway for
    				Selectively Exciting Solute Molecules to High Vibrational States via Solvent
    				Vibration-Polariton Pumping}},\ }\href
    	{https://doi.org/10.1038/s41467-022-31703-8} {\bibfield  {journal} {\bibinfo
    			{journal} {Nat. Commun.}\ }\textbf {\bibinfo {volume} {13}},\ \bibinfo
    		{pages} {4203} (\bibinfo {year} {2022})}\BibitemShut {NoStop}%
    	\bibitem [{\citenamefont {Flick}\ \emph {et~al.}(2017)\citenamefont {Flick},
    		\citenamefont {Ruggenthaler}, \citenamefont {Appel},\ and\ \citenamefont
    		{Rubio}}]{Flick2017}%
    	\BibitemOpen
    	\bibfield  {author} {\bibinfo {author} {\bibfnamefont {J.}~\bibnamefont
    			{Flick}}, \bibinfo {author} {\bibfnamefont {M.}~\bibnamefont {Ruggenthaler}},
    		\bibinfo {author} {\bibfnamefont {H.}~\bibnamefont {Appel}},\ and\ \bibinfo
    		{author} {\bibfnamefont {A.}~\bibnamefont {Rubio}},\ }\bibfield  {title}
    	{\bibinfo {title} {{Atoms and Molecules in Cavities, from Weak to Strong
    				Coupling in Quantum-Electrodynamics (QED) Chemistry}},\ }\href
    	{https://doi.org/10.1073/pnas.1615509114} {\bibfield  {journal} {\bibinfo
    			{journal} {Proc. Natl. Acad. Sci.}\ }\textbf {\bibinfo {volume} {114}},\
    		\bibinfo {pages} {3026} (\bibinfo {year} {2017})}\BibitemShut {NoStop}%
    	\bibitem [{\citenamefont {Sidler}\ \emph {et~al.}(2021)\citenamefont {Sidler},
    		\citenamefont {Sch{\"{a}}fer}, \citenamefont {Ruggenthaler},\ and\
    		\citenamefont {Rubio}}]{Sidler2021}%
    	\BibitemOpen
    	\bibfield  {author} {\bibinfo {author} {\bibfnamefont {D.}~\bibnamefont
    			{Sidler}}, \bibinfo {author} {\bibfnamefont {C.}~\bibnamefont
    			{Sch{\"{a}}fer}}, \bibinfo {author} {\bibfnamefont {M.}~\bibnamefont
    			{Ruggenthaler}},\ and\ \bibinfo {author} {\bibfnamefont {A.}~\bibnamefont
    			{Rubio}},\ }\bibfield  {title} {\bibinfo {title} {{Polaritonic Chemistry:
    				Collective Strong Coupling Implies Strong Local Modification of Chemical
    				Properties}},\ }\href {https://doi.org/10.1021/acs.jpclett.0c03436}
    	{\bibfield  {journal} {\bibinfo  {journal} {J. Phys. Chem. Lett.}\ }\textbf
    		{\bibinfo {volume} {12}},\ \bibinfo {pages} {508} (\bibinfo {year}
    		{2021})}\BibitemShut {NoStop}%
    	\bibitem [{\citenamefont {Yang}\ \emph {et~al.}(2021)\citenamefont {Yang},
    		\citenamefont {Ou}, \citenamefont {Pei}, \citenamefont {Wang}, \citenamefont
    		{Weng}, \citenamefont {Shuai}, \citenamefont {Mullen},\ and\ \citenamefont
    		{Shao}}]{Yang2021QEDFT}%
    	\BibitemOpen
    	\bibfield  {author} {\bibinfo {author} {\bibfnamefont {J.}~\bibnamefont
    			{Yang}}, \bibinfo {author} {\bibfnamefont {Q.}~\bibnamefont {Ou}}, \bibinfo
    		{author} {\bibfnamefont {Z.}~\bibnamefont {Pei}}, \bibinfo {author}
    		{\bibfnamefont {H.}~\bibnamefont {Wang}}, \bibinfo {author} {\bibfnamefont
    			{B.}~\bibnamefont {Weng}}, \bibinfo {author} {\bibfnamefont {Z.}~\bibnamefont
    			{Shuai}}, \bibinfo {author} {\bibfnamefont {K.}~\bibnamefont {Mullen}},\ and\
    		\bibinfo {author} {\bibfnamefont {Y.}~\bibnamefont {Shao}},\ }\bibfield
    	{title} {\bibinfo {title} {{Quantum-Electrodynamical Time-Dependent Density
    				Functional Theory within Gaussian Atomic Basis}},\ }\href
    	{https://doi.org/10.1063/5.0057542} {\bibfield  {journal} {\bibinfo
    			{journal} {J. Chem. Phys.}\ }\textbf {\bibinfo {volume} {155}},\ \bibinfo
    		{pages} {064107} (\bibinfo {year} {2021})}\BibitemShut {NoStop}%
    	\bibitem [{\citenamefont {Wei}\ \emph {et~al.}(2021)\citenamefont {Wei},
    		\citenamefont {Lee}, \citenamefont {Chou}, \citenamefont {Scholes},
    		\citenamefont {Schatz},\ and\ \citenamefont {Hsu}}]{Wei2021}%
    	\BibitemOpen
    	\bibfield  {author} {\bibinfo {author} {\bibfnamefont {Y.-C.}\ \bibnamefont
    			{Wei}}, \bibinfo {author} {\bibfnamefont {M.-W.}\ \bibnamefont {Lee}},
    		\bibinfo {author} {\bibfnamefont {P.-T.}\ \bibnamefont {Chou}}, \bibinfo
    		{author} {\bibfnamefont {G.~D.}\ \bibnamefont {Scholes}}, \bibinfo {author}
    		{\bibfnamefont {G.~C.}\ \bibnamefont {Schatz}},\ and\ \bibinfo {author}
    		{\bibfnamefont {L.-Y.}\ \bibnamefont {Hsu}},\ }\bibfield  {title} {\bibinfo
    		{title} {{Can Nanocavities Significantly Enhance Resonance Energy Transfer in
    				a Single Donor–Acceptor Pair?}},\ }\href
    	{https://doi.org/10.1021/acs.jpcc.1c04623} {\bibfield  {journal} {\bibinfo
    			{journal} {J. Phys. Chem. C}\ }\textbf {\bibinfo {volume} {125}},\ \bibinfo
    		{pages} {18119} (\bibinfo {year} {2021})}\BibitemShut {NoStop}%
    	\bibitem [{\citenamefont {{Terry Weatherly}}\ \emph {et~al.}(2023)\citenamefont
    		{{Terry Weatherly}}, \citenamefont {Provazza}, \citenamefont {Weiss},\ and\
    		\citenamefont {Tempelaar}}]{TerryWeatherly2023}%
    	\BibitemOpen
    	\bibfield  {author} {\bibinfo {author} {\bibfnamefont {C.~K.}\ \bibnamefont
    			{{Terry Weatherly}}}, \bibinfo {author} {\bibfnamefont {J.}~\bibnamefont
    			{Provazza}}, \bibinfo {author} {\bibfnamefont {E.~A.}\ \bibnamefont
    			{Weiss}},\ and\ \bibinfo {author} {\bibfnamefont {R.}~\bibnamefont
    			{Tempelaar}},\ }\bibfield  {title} {\bibinfo {title} {{Theory Predicts
    				UV/Vis-to-IR Photonic Down Conversion Mediated by Excited State Vibrational
    				Polaritons}},\ }\href {https://doi.org/10.1038/s41467-023-40400-z} {\bibfield
    		{journal} {\bibinfo  {journal} {Nat. Commun.}\ }\textbf {\bibinfo {volume}
    			{14}},\ \bibinfo {pages} {4804} (\bibinfo {year} {2023})}\BibitemShut
    	{NoStop}%
    	\bibitem [{\citenamefont {Li}\ and\ \citenamefont
    		{Hammes-Schiffer}(2023)}]{Li2023QMMM}%
    	\BibitemOpen
    	\bibfield  {author} {\bibinfo {author} {\bibfnamefont {T.~E.}\ \bibnamefont
    			{Li}}\ and\ \bibinfo {author} {\bibfnamefont {S.}~\bibnamefont
    			{Hammes-Schiffer}},\ }\bibfield  {title} {\bibinfo {title} {{QM/MM Modeling
    				of Vibrational Polariton Induced Energy Transfer and Chemical Dynamics}},\
    	}\href {https://doi.org/10.1021/JACS.2C10170/SUPPL_FILE/JA2C10170_SI_001.PDF}
    	{\bibfield  {journal} {\bibinfo  {journal} {J. Am. Chem. Soc.}\ }\textbf
    		{\bibinfo {volume} {145}},\ \bibinfo {pages} {377} (\bibinfo {year}
    		{2023})}\BibitemShut {NoStop}%
    	\bibitem [{\citenamefont {Wei}\ and\ \citenamefont {Hsu}(2023)}]{Wei2023}%
    	\BibitemOpen
    	\bibfield  {author} {\bibinfo {author} {\bibfnamefont {Y.-C.}\ \bibnamefont
    			{Wei}}\ and\ \bibinfo {author} {\bibfnamefont {L.-Y.}\ \bibnamefont {Hsu}},\
    	}\bibfield  {title} {\bibinfo {title} {{Polaritonic Huang–Rhys Factor:
    				Basic Concepts and Quantifying Light–Matter Interactions in Media}},\
    	}\href {https://doi.org/10.1021/acs.jpclett.3c00065} {\bibfield  {journal}
    		{\bibinfo  {journal} {J. Phys. Chem. Lett.}\ }\textbf {\bibinfo {volume}
    			{14}},\ \bibinfo {pages} {2395} (\bibinfo {year} {2023})}\BibitemShut
    	{NoStop}%
    	\bibitem [{\citenamefont {Fischer}\ and\ \citenamefont
    		{Saalfrank}(2023)}]{Fischer2023}%
    	\BibitemOpen
    	\bibfield  {author} {\bibinfo {author} {\bibfnamefont {E.~W.}\ \bibnamefont
    			{Fischer}}\ and\ \bibinfo {author} {\bibfnamefont {P.}~\bibnamefont
    			{Saalfrank}},\ }\bibfield  {title} {\bibinfo {title} {{Beyond Cavity
    				Born–Oppenheimer: On Nonadiabatic Coupling and Effective Ground State
    				Hamiltonians in Vibro-Polaritonic Chemistry}},\ }\bibfield  {journal}
    	{\bibinfo  {journal} {J. Chem. Theory Comput.}\ }\href
    	{https://doi.org/10.1021/acs.jctc.3c00708} {10.1021/acs.jctc.3c00708}
    	(\bibinfo {year} {2023})\BibitemShut {NoStop}%
    	\bibitem [{\citenamefont {Sukharev}\ \emph {et~al.}(2023)\citenamefont
    		{Sukharev}, \citenamefont {Subotnik},\ and\ \citenamefont
    		{Nitzan}}]{Sukharev2023}%
    	\BibitemOpen
    	\bibfield  {author} {\bibinfo {author} {\bibfnamefont {M.}~\bibnamefont
    			{Sukharev}}, \bibinfo {author} {\bibfnamefont {J.}~\bibnamefont {Subotnik}},\
    		and\ \bibinfo {author} {\bibfnamefont {A.}~\bibnamefont {Nitzan}},\
    	}\bibfield  {title} {\bibinfo {title} {{Dissociation Slowdown by Collective
    				Optical Response under Strong Coupling Conditions}},\ }\bibfield  {journal}
    	{\bibinfo  {journal} {J. Chem. Phys.}\ }\textbf {\bibinfo {volume} {158}},\
    	\href {https://doi.org/10.1063/5.0133972/2868724} {10.1063/5.0133972/2868724}
    	(\bibinfo {year} {2023})\BibitemShut {NoStop}%
    	\bibitem [{\citenamefont {Aroeira}\ \emph {et~al.}(2023)\citenamefont
    		{Aroeira}, \citenamefont {Kairys},\ and\ \citenamefont
    		{Ribeiro}}]{Aroeira2023}%
    	\BibitemOpen
    	\bibfield  {author} {\bibinfo {author} {\bibfnamefont {G.~J.~R.}\
    			\bibnamefont {Aroeira}}, \bibinfo {author} {\bibfnamefont {K.~T.}\
    			\bibnamefont {Kairys}},\ and\ \bibinfo {author} {\bibfnamefont {R.~F.}\
    			\bibnamefont {Ribeiro}},\ }\bibfield  {title} {\bibinfo {title} {{Theoretical
    				Analysis of Exciton Wave Packet Dynamics in Polaritonic Wires}},\ }\href
    	{https://doi.org/10.1021/acs.jpclett.3c01082} {\bibfield  {journal} {\bibinfo
    			{journal} {J. Phys. Chem. Lett.}\ }\textbf {\bibinfo {volume} {14}},\
    		\bibinfo {pages} {5681} (\bibinfo {year} {2023})},\ \Eprint
    	{https://arxiv.org/abs/2304.11453} {arXiv:2304.11453} \BibitemShut {NoStop}%
    	\bibitem [{\citenamefont {Moore}(1970)}]{Moore1970}%
    	\BibitemOpen
    	\bibfield  {author} {\bibinfo {author} {\bibfnamefont {G.~T.}\ \bibnamefont
    			{Moore}},\ }\bibfield  {title} {\bibinfo {title} {{Quantum Theory of the
    				Electromagnetic Field in a Variable-Length One-Dimensional Cavity}},\ }\href
    	{https://doi.org/10.1063/1.1665432} {\bibfield  {journal} {\bibinfo
    			{journal} {J. Math. Phys.}\ }\textbf {\bibinfo {volume} {11}},\ \bibinfo
    		{pages} {2679} (\bibinfo {year} {1970})}\BibitemShut {NoStop}%
    	\bibitem [{\citenamefont {Schwinger}(1993)}]{Schwinger1993}%
    	\BibitemOpen
    	\bibfield  {author} {\bibinfo {author} {\bibfnamefont {J.}~\bibnamefont
    			{Schwinger}},\ }\bibfield  {title} {\bibinfo {title} {{Casimir Light: The
    				Source.}},\ }\href {https://doi.org/10.1073/pnas.90.6.2105} {\bibfield
    		{journal} {\bibinfo  {journal} {Proc. Natl. Acad. Sci.}\ }\textbf {\bibinfo
    			{volume} {90}},\ \bibinfo {pages} {2105} (\bibinfo {year}
    		{1993})}\BibitemShut {NoStop}%
    	\bibitem [{\citenamefont {Uhlmann}\ \emph {et~al.}(2004)\citenamefont
    		{Uhlmann}, \citenamefont {Plunien}, \citenamefont {Sch{\"{u}}tzhold},\ and\
    		\citenamefont {Soff}}]{Uhlmann2004}%
    	\BibitemOpen
    	\bibfield  {author} {\bibinfo {author} {\bibfnamefont {M.}~\bibnamefont
    			{Uhlmann}}, \bibinfo {author} {\bibfnamefont {G.}~\bibnamefont {Plunien}},
    		\bibinfo {author} {\bibfnamefont {R.}~\bibnamefont {Sch{\"{u}}tzhold}},\ and\
    		\bibinfo {author} {\bibfnamefont {G.}~\bibnamefont {Soff}},\ }\bibfield
    	{title} {\bibinfo {title} {{Resonant Cavity Photon Creation via the Dynamical
    				Casimir Effect}},\ }\href {https://doi.org/10.1103/PhysRevLett.93.193601}
    	{\bibfield  {journal} {\bibinfo  {journal} {Phys. Rev. Lett.}\ }\textbf
    		{\bibinfo {volume} {93}},\ \bibinfo {pages} {193601} (\bibinfo {year}
    		{2004})}\BibitemShut {NoStop}%
    	\bibitem [{\citenamefont {Dodonov}(2010)}]{Dodonov2010}%
    	\BibitemOpen
    	\bibfield  {author} {\bibinfo {author} {\bibfnamefont {V.~V.}\ \bibnamefont
    			{Dodonov}},\ }\bibfield  {title} {\bibinfo {title} {{Current Status of the
    				Dynamical Casimir Effect}},\ }\href
    	{https://doi.org/10.1088/0031-8949/82/03/038105} {\bibfield  {journal}
    		{\bibinfo  {journal} {Phys. Scr.}\ }\textbf {\bibinfo {volume} {82}},\
    		\bibinfo {pages} {038105} (\bibinfo {year} {2010})}\BibitemShut {NoStop}%
    	\bibitem [{\citenamefont {Macr{\`{i}}}\ \emph {et~al.}(2018)\citenamefont
    		{Macr{\`{i}}}, \citenamefont {Ridolfo}, \citenamefont {{Di Stefano}},
    		\citenamefont {Kockum}, \citenamefont {Nori},\ and\ \citenamefont
    		{Savasta}}]{Macri2018}%
    	\BibitemOpen
    	\bibfield  {author} {\bibinfo {author} {\bibfnamefont {V.}~\bibnamefont
    			{Macr{\`{i}}}}, \bibinfo {author} {\bibfnamefont {A.}~\bibnamefont
    			{Ridolfo}}, \bibinfo {author} {\bibfnamefont {O.}~\bibnamefont {{Di
    					Stefano}}}, \bibinfo {author} {\bibfnamefont {A.~F.}\ \bibnamefont {Kockum}},
    		\bibinfo {author} {\bibfnamefont {F.}~\bibnamefont {Nori}},\ and\ \bibinfo
    		{author} {\bibfnamefont {S.}~\bibnamefont {Savasta}},\ }\bibfield  {title}
    	{\bibinfo {title} {{Nonperturbative Dynamical Casimir Effect in
    				Optomechanical Systems: Vacuum Casimir-Rabi Splittings}},\ }\href
    	{https://doi.org/10.1103/PHYSREVX.8.011031/FIGURES/9/MEDIUM} {\bibfield
    		{journal} {\bibinfo  {journal} {Phys. Rev. X}\ }\textbf {\bibinfo {volume}
    			{8}},\ \bibinfo {pages} {011031} (\bibinfo {year} {2018})}\BibitemShut
    	{NoStop}%
    	\bibitem [{\citenamefont {Li}\ \emph {et~al.}(2005)\citenamefont {Li},
    		\citenamefont {Tully}, \citenamefont {Schlegel},\ and\ \citenamefont
    		{Frisch}}]{Li2005Eh}%
    	\BibitemOpen
    	\bibfield  {author} {\bibinfo {author} {\bibfnamefont {X.}~\bibnamefont
    			{Li}}, \bibinfo {author} {\bibfnamefont {J.~C.}\ \bibnamefont {Tully}},
    		\bibinfo {author} {\bibfnamefont {H.~B.}\ \bibnamefont {Schlegel}},\ and\
    		\bibinfo {author} {\bibfnamefont {M.~J.}\ \bibnamefont {Frisch}},\ }\bibfield
    	{title} {\bibinfo {title} {{Ab initio Ehrenfest dynamics}},\ }\href
    	{https://doi.org/10.1063/1.2008258} {\bibfield  {journal} {\bibinfo
    			{journal} {J. Chem. Phys.}\ }\textbf {\bibinfo {volume} {123}},\ \bibinfo
    		{pages} {084106} (\bibinfo {year} {2005})}\BibitemShut {NoStop}%
    	\bibitem [{\citenamefont {Miller}(1978)}]{Miller1978}%
    	\BibitemOpen
    	\bibfield  {author} {\bibinfo {author} {\bibfnamefont {W.~H.}\ \bibnamefont
    			{Miller}},\ }\bibfield  {title} {\bibinfo {title} {{A Classical/Semiclassical
    				Theory for the Interaction of Infrared Radiation with Molecular Systems}},\
    	}\href {https://doi.org/10.1063/1.436793} {\bibfield  {journal} {\bibinfo
    			{journal} {J. Chem. Phys.}\ }\textbf {\bibinfo {volume} {69}},\ \bibinfo
    		{pages} {2188} (\bibinfo {year} {1978})}\BibitemShut {NoStop}%
    	\bibitem [{\citenamefont {Li}\ \emph {et~al.}(2018)\citenamefont {Li},
    		\citenamefont {Nitzan}, \citenamefont {Sukharev}, \citenamefont {Martinez},
    		\citenamefont {Chen},\ and\ \citenamefont {Subotnik}}]{Li2018Spontaneous}%
    	\BibitemOpen
    	\bibfield  {author} {\bibinfo {author} {\bibfnamefont {T.~E.}\ \bibnamefont
    			{Li}}, \bibinfo {author} {\bibfnamefont {A.}~\bibnamefont {Nitzan}}, \bibinfo
    		{author} {\bibfnamefont {M.}~\bibnamefont {Sukharev}}, \bibinfo {author}
    		{\bibfnamefont {T.}~\bibnamefont {Martinez}}, \bibinfo {author}
    		{\bibfnamefont {H.-T.}\ \bibnamefont {Chen}},\ and\ \bibinfo {author}
    		{\bibfnamefont {J.~E.}\ \bibnamefont {Subotnik}},\ }\bibfield  {title}
    	{\bibinfo {title} {{Mixed Quantum--Classical Electrodynamics: Understanding
    				Spontaneous Decay and Zero--Point Energy}},\ }\href
    	{https://doi.org/10.1103/PhysRevA.97.032105} {\bibfield  {journal} {\bibinfo
    			{journal} {Phys. Rev. A}\ }\textbf {\bibinfo {volume} {97}},\ \bibinfo
    		{pages} {032105} (\bibinfo {year} {2018})}\BibitemShut {NoStop}%
    	\bibitem [{\citenamefont {Hoffmann}\ \emph {et~al.}(2018)\citenamefont
    		{Hoffmann}, \citenamefont {Appel}, \citenamefont {Rubio},\ and\ \citenamefont
    		{Maitra}}]{Hoffmann2018}%
    	\BibitemOpen
    	\bibfield  {author} {\bibinfo {author} {\bibfnamefont {N.~M.}\ \bibnamefont
    			{Hoffmann}}, \bibinfo {author} {\bibfnamefont {H.}~\bibnamefont {Appel}},
    		\bibinfo {author} {\bibfnamefont {A.}~\bibnamefont {Rubio}},\ and\ \bibinfo
    		{author} {\bibfnamefont {N.~T.}\ \bibnamefont {Maitra}},\ }\bibfield  {title}
    	{\bibinfo {title} {{Light-matter Interactions via the Exact Factorization
    				Approach}},\ }\href {https://doi.org/10.1140/epjb/e2018-90177-6} {\bibfield
    		{journal} {\bibinfo  {journal} {Eur. Phys. J. B}\ }\textbf {\bibinfo {volume}
    			{91}},\ \bibinfo {pages} {180} (\bibinfo {year} {2018})}\BibitemShut
    	{NoStop}%
    	\bibitem [{\citenamefont {Sch{\"{a}}fer}\ \emph {et~al.}(2020)\citenamefont
    		{Sch{\"{a}}fer}, \citenamefont {Ruggenthaler}, \citenamefont {Rokaj},\ and\
    		\citenamefont {Rubio}}]{Schafer2020}%
    	\BibitemOpen
    	\bibfield  {author} {\bibinfo {author} {\bibfnamefont {C.}~\bibnamefont
    			{Sch{\"{a}}fer}}, \bibinfo {author} {\bibfnamefont {M.}~\bibnamefont
    			{Ruggenthaler}}, \bibinfo {author} {\bibfnamefont {V.}~\bibnamefont
    			{Rokaj}},\ and\ \bibinfo {author} {\bibfnamefont {A.}~\bibnamefont {Rubio}},\
    	}\bibfield  {title} {\bibinfo {title} {{Relevance of the Quadratic
    				Diamagnetic and Self-Polarization Terms in Cavity Quantum Electrodynamics}},\
    	}\href {https://doi.org/10.1021/acsphotonics.9b01649} {\bibfield  {journal}
    		{\bibinfo  {journal} {ACS Photonics}\ }\textbf {\bibinfo {volume} {7}},\
    		\bibinfo {pages} {975} (\bibinfo {year} {2020})}\BibitemShut {NoStop}%
    	\bibitem [{\citenamefont {Taylor}\ \emph {et~al.}(2020)\citenamefont {Taylor},
    		\citenamefont {Mandal}, \citenamefont {Zhou},\ and\ \citenamefont
    		{Huo}}]{Taylor2020}%
    	\BibitemOpen
    	\bibfield  {author} {\bibinfo {author} {\bibfnamefont {M.~A.~D.}\
    			\bibnamefont {Taylor}}, \bibinfo {author} {\bibfnamefont {A.}~\bibnamefont
    			{Mandal}}, \bibinfo {author} {\bibfnamefont {W.}~\bibnamefont {Zhou}},\ and\
    		\bibinfo {author} {\bibfnamefont {P.}~\bibnamefont {Huo}},\ }\bibfield
    	{title} {\bibinfo {title} {{Resolution of Gauge Ambiguities in Molecular
    				Cavity Quantum Electrodynamics}},\ }\href
    	{https://doi.org/10.1103/PhysRevLett.125.123602} {\bibfield  {journal}
    		{\bibinfo  {journal} {Phys. Rev. Lett.}\ }\textbf {\bibinfo {volume} {125}},\
    		\bibinfo {pages} {123602} (\bibinfo {year} {2020})}\BibitemShut {NoStop}%
    	\bibitem [{\citenamefont {Stokes}\ and\ \citenamefont
    		{Nazir}(2022)}]{Stokes2022}%
    	\BibitemOpen
    	\bibfield  {author} {\bibinfo {author} {\bibfnamefont {A.}~\bibnamefont
    			{Stokes}}\ and\ \bibinfo {author} {\bibfnamefont {A.}~\bibnamefont {Nazir}},\
    	}\bibfield  {title} {\bibinfo {title} {{Implications of Gauge Freedom for
    				Nonrelativistic Quantum Electrodynamics}},\ }\href
    	{https://doi.org/10.1103/REVMODPHYS.94.045003/FIGURES/13/MEDIUM} {\bibfield
    		{journal} {\bibinfo  {journal} {Rev. Mod.Phys.}\ }\textbf {\bibinfo {volume}
    			{94}},\ \bibinfo {pages} {045003} (\bibinfo {year} {2022})},\ \Eprint
    	{https://arxiv.org/abs/2009.10662} {arXiv:2009.10662} \BibitemShut {NoStop}%
    	\bibitem [{Note1()}]{Note1}%
    	\BibitemOpen
    	\bibinfo {note} {The classical photonic energy is calculated by $\protect
    		\frac {1}{2}p_{\protect \rm c}^2 + \protect \frac {1}{2}\omega _{\protect \rm
    			c}^2(q_{\protect \rm c} + \lambda _{\protect \rm c} \left \langle \protect
    		\cc@accent {"705E}{\mu }\right \rangle /\omega _{\protect \rm c} )^2$ instead
    		of $\protect \frac {1}{2}p_{\protect \rm c}^2 + \omega _{\protect \rm
    			c}\lambda _{\protect \rm c} q_{\protect \rm c}\left \langle \protect
    		\cc@accent {"705E}{\mu }\right \rangle + \protect \frac {1}{2}\lambda
    		_{\protect \rm c}^2 \left \langle \protect \cc@accent {"705E}{\mu }^2\right
    		\rangle $ due to a recent suggestion \cite {Taylor2020} considering gauge
    		invariance.}\BibitemShut {Stop}%
    	\bibitem [{Note2()}]{Note2}%
    	\BibitemOpen
    	\bibinfo {note} {Here no dissipation pathway is considered}\BibitemShut
    	{NoStop}%
    	\bibitem [{\citenamefont {Xiang}\ \emph {et~al.}(2019)\citenamefont {Xiang},
    		\citenamefont {Ribeiro}, \citenamefont {Chen}, \citenamefont {Wang},
    		\citenamefont {Du}, \citenamefont {Yuen-Zhou},\ and\ \citenamefont
    		{Xiong}}]{Xiang2019State}%
    	\BibitemOpen
    	\bibfield  {author} {\bibinfo {author} {\bibfnamefont {B.}~\bibnamefont
    			{Xiang}}, \bibinfo {author} {\bibfnamefont {R.~F.}\ \bibnamefont {Ribeiro}},
    		\bibinfo {author} {\bibfnamefont {L.}~\bibnamefont {Chen}}, \bibinfo {author}
    		{\bibfnamefont {J.}~\bibnamefont {Wang}}, \bibinfo {author} {\bibfnamefont
    			{M.}~\bibnamefont {Du}}, \bibinfo {author} {\bibfnamefont {J.}~\bibnamefont
    			{Yuen-Zhou}},\ and\ \bibinfo {author} {\bibfnamefont {W.}~\bibnamefont
    			{Xiong}},\ }\bibfield  {title} {\bibinfo {title} {{State-Selective Polariton
    				to Dark State Relaxation Dynamics}},\ }\href
    	{https://doi.org/10.1021/acs.jpca.9b04601} {\bibfield  {journal} {\bibinfo
    			{journal} {J. Phys. Chem. A}\ }\textbf {\bibinfo {volume} {123}},\ \bibinfo
    		{pages} {5918} (\bibinfo {year} {2019})}\BibitemShut {NoStop}%
    	\bibitem [{\citenamefont {Li}\ \emph {et~al.}(2021{\natexlab{b}})\citenamefont
    		{Li}, \citenamefont {Nitzan},\ and\ \citenamefont
    		{Subotnik}}]{Li2020Nonlinear}%
    	\BibitemOpen
    	\bibfield  {author} {\bibinfo {author} {\bibfnamefont {T.~E.}\ \bibnamefont
    			{Li}}, \bibinfo {author} {\bibfnamefont {A.}~\bibnamefont {Nitzan}},\ and\
    		\bibinfo {author} {\bibfnamefont {J.~E.}\ \bibnamefont {Subotnik}},\
    	}\bibfield  {title} {\bibinfo {title} {{Cavity Molecular Dynamics Simulations
    				of Vibrational Polariton-Enhanced Molecular Nonlinear Absorption}},\ }\href
    	{https://doi.org/10.1063/5.0037623} {\bibfield  {journal} {\bibinfo
    			{journal} {J. Chem. Phys.}\ }\textbf {\bibinfo {volume} {154}},\ \bibinfo
    		{pages} {094124} (\bibinfo {year} {2021}{\natexlab{b}})}\BibitemShut
    	{NoStop}%
    	\bibitem [{\citenamefont {Hammes-Schiffer}(2015)}]{Hammes-Schiffer2015}%
    	\BibitemOpen
    	\bibfield  {author} {\bibinfo {author} {\bibfnamefont {S.}~\bibnamefont
    			{Hammes-Schiffer}},\ }\bibfield  {title} {\bibinfo {title} {{Proton-Coupled
    				Electron Transfer: Moving Together and Charging Forward}},\ }\href
    	{https://doi.org/10.1021/jacs.5b04087} {\bibfield  {journal} {\bibinfo
    			{journal} {J. Am. Chem. Soc.}\ }\textbf {\bibinfo {volume} {137}},\ \bibinfo
    		{pages} {8860} (\bibinfo {year} {2015})}\BibitemShut {NoStop}%
    	\bibitem [{\citenamefont {{\L}empicka-Mirek}\ \emph {et~al.}(2022)\citenamefont
    		{{\L}empicka-Mirek}, \citenamefont {Kr{\'{o}}l}, \citenamefont {Sigurdsson},
    		\citenamefont {Wincukiewicz}, \citenamefont {Morawiak}, \citenamefont
    		{Mazur}, \citenamefont {Muszy{\'{n}}ski}, \citenamefont {Piecek},
    		\citenamefont {Kula}, \citenamefont {Stefaniuk}, \citenamefont
    		{Kami{\'{n}}ska}, \citenamefont {{De Marco}}, \citenamefont {Lagoudakis},
    		\citenamefont {Ballarini}, \citenamefont {Sanvitto}, \citenamefont
    		{Szczytko},\ and\ \citenamefont {Pi{\c{e}}tka}}]{empicka-Mirek2022}%
    	\BibitemOpen
    	\bibfield  {author} {\bibinfo {author} {\bibfnamefont {K.}~\bibnamefont
    			{{\L}empicka-Mirek}}, \bibinfo {author} {\bibfnamefont {M.}~\bibnamefont
    			{Kr{\'{o}}l}}, \bibinfo {author} {\bibfnamefont {H.}~\bibnamefont
    			{Sigurdsson}}, \bibinfo {author} {\bibfnamefont {A.}~\bibnamefont
    			{Wincukiewicz}}, \bibinfo {author} {\bibfnamefont {P.}~\bibnamefont
    			{Morawiak}}, \bibinfo {author} {\bibfnamefont {R.}~\bibnamefont {Mazur}},
    		\bibinfo {author} {\bibfnamefont {M.}~\bibnamefont {Muszy{\'{n}}ski}},
    		\bibinfo {author} {\bibfnamefont {W.}~\bibnamefont {Piecek}}, \bibinfo
    		{author} {\bibfnamefont {P.}~\bibnamefont {Kula}}, \bibinfo {author}
    		{\bibfnamefont {T.}~\bibnamefont {Stefaniuk}}, \bibinfo {author}
    		{\bibfnamefont {M.}~\bibnamefont {Kami{\'{n}}ska}}, \bibinfo {author}
    		{\bibfnamefont {L.}~\bibnamefont {{De Marco}}}, \bibinfo {author}
    		{\bibfnamefont {P.~G.}\ \bibnamefont {Lagoudakis}}, \bibinfo {author}
    		{\bibfnamefont {D.}~\bibnamefont {Ballarini}}, \bibinfo {author}
    		{\bibfnamefont {D.}~\bibnamefont {Sanvitto}}, \bibinfo {author}
    		{\bibfnamefont {J.}~\bibnamefont {Szczytko}},\ and\ \bibinfo {author}
    		{\bibfnamefont {B.}~\bibnamefont {Pi{\c{e}}tka}},\ }\bibfield  {title}
    	{\bibinfo {title} {{Electrically tunable Berry curvature and strong
    				light-matter coupling in liquid crystal microcavities with 2D perovskite}},\
    	}\bibfield  {journal} {\bibinfo  {journal} {Sci. Adv.}\ }\textbf {\bibinfo
    		{volume} {8}},\ \href {https://doi.org/10.1126/sciadv.abq7533}
    	{10.1126/sciadv.abq7533} (\bibinfo {year} {2022})\BibitemShut {NoStop}%
    	\bibitem [{\citenamefont {Stemo}\ \emph {et~al.}(2022)\citenamefont {Stemo},
    		\citenamefont {Yamada}, \citenamefont {Katsuki},\ and\ \citenamefont
    		{Yanagi}}]{Stemo2022}%
    	\BibitemOpen
    	\bibfield  {author} {\bibinfo {author} {\bibfnamefont {G.}~\bibnamefont
    			{Stemo}}, \bibinfo {author} {\bibfnamefont {H.}~\bibnamefont {Yamada}},
    		\bibinfo {author} {\bibfnamefont {H.}~\bibnamefont {Katsuki}},\ and\ \bibinfo
    		{author} {\bibfnamefont {H.}~\bibnamefont {Yanagi}},\ }\bibfield  {title}
    	{\bibinfo {title} {{Influence of Vibrational Strong Coupling on an Ordered
    				Liquid Crystal}},\ }\href {https://doi.org/10.1021/acs.jpcb.2c04004}
    	{\bibfield  {journal} {\bibinfo  {journal} {J. Phys. Chem. B}\ }\textbf
    		{\bibinfo {volume} {126}},\ \bibinfo {pages} {9399} (\bibinfo {year}
    		{2022})}\BibitemShut {NoStop}%
    \end{thebibliography}

    %

    \end{document}


\title{Supporting Information \\ \textit{ } \\ Theory of Supervibronic Transitions via Casimir Polaritons}
	
	\author{Tao E. Li}%
	\email{taoeli@udel.edu}
	\affiliation{Department of Physics and Astronomy, University of Delaware, Newark, Delaware 19716, USA}
		
	\setcounter{equation}{0}
	\setcounter{figure}{0}
	\setcounter{table}{0}
	\renewcommand{\theequation}{S\arabic{equation}}
	\renewcommand{\thefigure}{S\arabic{figure}}
	\renewcommand{\bibnumfmt}[1]{[S#1]}
	\renewcommand{\citenumfont}[1]{S#1}
	\renewcommand{\thepage}{S\arabic{page}}
	
	\maketitle
	

 \section{Details of numerical simulations}
    
    With the cavity Ehrenfest approximation, we may rewrite the molecular Hamiltonian in Eq. (2) as 
    \begin{equation}\label{eq:HM_model_sc}
        \hH_{\rm M} = \omega_{\rm e} \sum_{i=1}^{N_{\rm e}} \hat{\sigma}_{+}^{i}\hat{\sigma}_{-}^{i} + \sum_{j=1}^{N_{\rm v}} \left ( \frac{p_j^2}{2} + \frac{1}{2}\omega_{\rm v}^2 q_j^2 \right ),
    \end{equation}
    where the classical momentum and position variables $p_j$ and $q_j$ have been used to replace $\hat{b}_j$ and $\hat{b}_{j}^{\dagger}$ with the well-known relations $\hat{b} = \frac{1}{\sqrt{2\omega}} (\omega \hat{q} + i\hat{p})$ and $\hat{b}^{\dagger} = \frac{1}{\sqrt{2\omega}} (\omega \hat{q} - i\hat{p})$. Similarly, the quantum vibrational dipole operator in Eq. (4) can also be replaced by the classical variable
    \begin{equation}\label{eq:dipole_v_classical}
        \mu_{\rm v} = \sum_{j=1}^{N_{\rm v}} d_{\rm v} q_{j}.
    \end{equation}
    The coupled electron-vibration-photon system is thus governed by the following semiclassical Pauli--Fierz Hamiltonian:
    \begin{equation}\label{eq:H_PF_semiclassical}
        \hH_{\rm PF} = \hH_{\rm M} + \frac{1}{2} p_{\rm c}^2 + \frac{1}{2}\omega_{\rm c}^2 \left ( q_{\rm c} + \frac{\lambda_{\rm c}}{\omega_{\rm c}}  \hat{\mu} \right  )^2,
    \end{equation}
    where $\hat{\mu} = \mu_{\rm v} + \hat{\mu}_e$, and $\hat{\mu}_e$ and $\mu_{\rm v}$ have been defined in Eqs. (3) and \eqref{eq:dipole_v_classical}, respectively. The corresponding energy expectation value of Eq. \eqref{eq:H_PF_semiclassical} is
    \begin{equation}\label{eq:H_PF_semiclassical_avg}
        \avg{\hH_{\rm PF}} = \avg{\hH_{\rm M}} + \frac{1}{2} p_{\rm c}^2 + \frac{1}{2}\omega_{\rm c}^2 \left ( q_{\rm c} + \frac{\lambda_{\rm c}}{\omega_{\rm c}}  \avg{\hat{\mu}} \right  )^2.
    \end{equation}
    When evaluating the energy expectation value of the dipole self-energy term (i.e., the term containing $\hat{\mu}^2$ in Eq. \eqref{eq:H_PF_semiclassical}), we use $\avg{\hat{\mu}}^2$ instead of $\avg{\hat{\mu}^2}$. Similar suggestions have been made in a recent work \cite{Taylor2020} discussing gauge invariance in molecular quantum electrodynamics. 
    
    According to Eqs. \eqref{eq:H_PF_semiclassical} and \eqref{eq:H_PF_semiclassical_avg}, the corresponding semiclassical equations of motion become:
    \begin{subequations}\label{eq:EOM_sc_original}
        \begin{align}
        \label{eq:EOM_sc_original-e}
            \frac{\partial}{\partial t} \ket{\psi_{\rm e}} &= -i \hH_{\rm sc} \ket{\psi_{\rm e}}, \\
            \label{eq:EOM_sc_original-c}
            \ddot{q}_{\rm c} &= -\omega_{\rm c}^2 q_{\rm c} - \omega_{\rm c}\lambda_{\rm c} \avg{\hat{\mu}},  \\
            \label{eq:EOM_sc_original-v}
            \ddot{q}_{j} &= -\omega_{\rm v}^2 q_{j} - \omega_{\rm c}\lambda_{\rm c} q_{\rm c} \frac{\partial \avg{\hat{\mu}}}{\partial q_{j}} 
            - \lambda_{\rm c}^2 \avg{\hat{\mu}} \frac{\partial \avg{\hat{\mu}}}{\partial q_{j}} .
        \end{align}
    \end{subequations}
    In Eq. \eqref{eq:EOM_sc_original-e}, $\ket{\psi_{\rm e}}$ denotes the wavefunction for the whole electronic subsystem, and 
    \begin{equation}\label{eq:Hsc}
        \hH_{\rm sc} = \omega_{\rm e} \sum_{i=1}^{N_{\rm e}} \hat{\sigma}_{+}^{i}\hat{\sigma}_{-}^{i} + \omega_{\rm c}\lambda_{\rm c}q_{\rm c} \hat{\mu} + \frac{1}{2}\lambda_{\rm c}^2 \hat{\mu}^2
    \end{equation}
    is Eq. \eqref{eq:H_PF_semiclassical} excluding scalar terms. In Eq. \eqref{eq:EOM_sc_original-c}, the classical cavity photon interacts with the mean-field value of the total dipole moment averaged over different electronic  states: $\avg{\hat{\mu}} = \mu_{\rm v} + \avgt{\psi_{\rm e}}{\hat{\mu}_{\rm e}}{\psi_{\rm e}}$. In Eq. \eqref{eq:EOM_sc_original-v}, $\partial \avg{\hat{\mu}}/\partial q_j = d_{\rm  v}$ according to Eq. \eqref{eq:dipole_v_classical}.

    In the collective regime, because both $N_{\rm e}$ and $N_{\rm v}$ can be macroscopically large, a direct propagation of Eq. \eqref{eq:EOM_sc_original} is computationally demanding. Instead, due to the symmetry of the system, we may efficiently simulate the above many-body dynamics with much fewer degrees of freedom. For example, in Eq.  \eqref{eq:EOM_sc_original-e}, assuming working under the symmetric Dicke superradiant states, we may use a density matrix of a single TLS $\hrho_{\rm e,s}$ to represent the average state of $N_{\rm e}$-TLS wavefunction $\ket{\psi_{\rm e}}$ by taking the partial trace over $N_{\rm e}-1$ TLSs: $\hrho_{\rm e,s} = \text{Tr}_{N_{\rm e}-1}\left ( {\ket{\psi_{\rm e}}\bra{\psi_{\rm e}}} \right )$; see also Ref. \cite{Breuer2007} on the use of an effective single-TLS Hamiltonian to describe the Dicke's superradiance process. Likewise, we can also obtain the single-molecule semiclassical Hamiltonian $\hH_{\rm sc,s}$ by taking the partial trace over $N_{\rm e}-1$ TLSs, i.e., $\hH_{\rm sc,s} = \avg{\hH_{\rm sc}}_{N_{\rm e}-1}$. Using the definition of $\hH_{\rm sc}$ in Eq. \eqref{eq:Hsc}, we obtain 
    \begin{equation}\label{eq:He_semi_single}
        \hH_{\rm sc,s} = \omega_{\rm e} \hat{\sigma}_{+} \hat{\sigma}_{-}  + \omega_{\rm c}\lambda_{\rm c}q_{\rm c} \hat{\mu}_{\rm e,s} + \frac{1}{2}\lambda_{\rm c}^2 \left [\hat{\mu}_{\rm e,s}^2 + 2(N_{\rm e}-1)\avg{\hat{\mu}_{\rm e,s}}\hat{\mu}_{\rm e,s} + 2\mu_{\rm v}\hat{\mu}_{\rm e,s}\right ] ,
    \end{equation}
    where $\hat{\mu}_{\rm e,s} = d_{\rm eg}( \hat{\sigma}_{+} + \hat{\sigma}_{-} ) + \left( \bar{d} \hat{1} + \frac{\Delta d}{2} \hat{\sigma}_z \right)$ denotes the single-molecule electronic dipole operator and $\avg{\hat{\mu}_{\rm e,s}} = \tr{\rho_{\rm e,s} \hat{\mu}_{\rm e,s}}$.  Hence, in the single-TLS basis, with the assumption that the electronic dynamics are evolved under the symmetric Dicke superradiant states, the electronic equations of motion become
    \begin{equation}\label{eq:drho_sdt}
        \frac{\partial}{\partial t} \hrho_{\rm e,s} = -i \left [ \hH_{\rm sc,s},\ \hrho_{\rm e,s} \right ] .
    \end{equation}
    Moreover, for the vibrational dynamics in Eq. \eqref{eq:EOM_sc_original-v}, the vibrational variables can be separated to one bright state $q_{\rm B} = \sum_{j} q_j / \sqrt{N_{\rm v}}$ (which is coupled to the cavity mode) and $N_{\rm v} - 1$ asymmetric dark states $q_{\text{D},k} = \sum_{j} \exp(2\pi j k/N_{\rm v}) q_j / \sqrt{N_{\rm v}}$ for $k = 1, 2, \cdots, N_{\rm v}$ (which are decoupled from the cavity mode). With this collective variable transformation, Eq. \eqref{eq:EOM_sc_original-v} becomes
    \begin{subequations}\label{eq:dbddt}
        \begin{align}
            \ddot{q}_{\rm B} &= -\omega_{\rm v}^2 q_{\rm B} - \omega_{\rm c}\lambda_{\rm c} q_{\rm c} \frac{\partial \avg{\hat{\mu}}}{\partial q_{\rm B}} 
            - \lambda_{\rm c}^2 \avg{\hat{\mu}} \frac{\partial \avg{\hat{\mu}}}{\partial q_{\rm B}} , \\
            \ddot{q}_{\text{D},k} &=  -\omega_{\rm v}^2 q_{\text{D},k} .
        \end{align}
    \end{subequations}
    where $\avg{\hat{\mu}} = \sqrt{N_{\rm v}}d_{\rm v} q_{\rm B} + N_{\rm e}\tr{\hrho_{\rm e,s}\hat{\mu}_{e,s}}$ and $\partial \avg{\hat{\mu}} / \partial q_{\rm B} = \sqrt{N_{\rm v}}d_{\rm v}$.
    
    With Eqs. \eqref{eq:drho_sdt} and \eqref{eq:dbddt}, we can replace the equations of motion in Eq. \eqref{eq:EOM_sc_original} by:
    \begin{subequations}\label{eq:EOM_sc_simplified}
        \begin{align}
            \frac{\partial}{\partial t} \hrho_{\rm e,s} &= -i \left [ \hH_{\rm sc,s},\ \hrho_{\rm e,s} \right ]
            \label{eq:EOM_sub_e-nodamping} ,
            \\
            \ddot{q}_{\rm c} &= -\omega_{\rm c}^2 q_{\rm c} - \omega_{\rm c}\lambda_{\rm c} \avg{\hat{\mu}} \label{eq:EOM_sub_ph-nodamping} ,\\
            \ddot{q}_{\rm B} &= -\omega_{\rm v}^2 q_{\rm B} - \omega_{\rm c}\lambda_{\rm c} q_{\rm c} \frac{\partial \avg{\hat{\mu}}}{\partial q_{\rm B}} 
            - \lambda_{\rm c}^2 \avg{\hat{\mu}} \frac{\partial \avg{\hat{\mu}}}{\partial q_{\rm B}}  \label{eq:EOM_sub_B-nodamping} ,\\
            \ddot{q}_{\text{D},k} &=  -\omega_{\rm v}^2 q_{\text{D},k} .
            \label{eq:EOM_sub_D-nodamping}
        \end{align}
    \end{subequations}
    At this moment, Eq. \eqref{eq:EOM_sc_simplified} does not contain any relaxation pathway. In reality, the electronic subsystem is not always evolved under the symmetric Dicke superradiant states and its relaxation and decoherence should be taken into account; due to the imperfectness of the cavity mirrors, the cavity photon has a lifetime from hundreds of fs to a few ps  (for VSC systems) \cite{Xiang2019State,Grafton2020,Li2020Nonlinear}; as far as the dynamics of the bright and the dark modes are concerned, it is well studied that the bright mode (or vibrational polaritons) can quickly dephase to the dark modes with a lifetime of a few ps \cite{Xiang2019State}.

    Taking these relaxation pathways into account, we can add a few phenomenological terms on top of  Eq. \eqref{eq:EOM_sc_simplified}:
    \begin{subequations}\label{eq:EOM_final_original}
        \begin{align}
            \frac{\partial}{\partial t} \hrho_{\rm e,s} &= -i \left [ \hH_{\rm sc,s},\ \hrho_{\rm e,s} \right ] + \hat{\mathcal{L}}_{\rm dmp}[\hat{\rho}_{\rm e,s}] + \hat{\mathcal{L}}_{\rm drv}[\hat{\rho}_{\rm e,s}]
            \label{eq:EOM_sub_e}
            ,\\
            \ddot{q}_{\rm c} &= -\omega_{\rm c}^2 q_{\rm c} - \omega_{\rm c}\lambda_{\rm c} \avg{\hat{\mu}} - \gamma_{\rm c} p_{\rm c} \label{eq:EOM_sub_ph} ,\\
            \ddot{q}_{\rm B} &= -\omega_{\rm v}^2 q_{\rm B} - \omega_{\rm c}\lambda_{\rm c} q_{\rm c} \frac{\partial \avg{\hat{\mu}}}{\partial q_{\rm B}} 
            - \lambda_{\rm c}^2 \avg{\hat{\mu}} \frac{\partial \avg{\hat{\mu}}}{\partial q_{\rm B}} - \gamma_{\rm v} \sum_{k}q_{\text{D},k}  \label{eq:EOM_sub_B} ,\\
            \ddot{q}_{\text{D},k} &=  -\omega_{\text{D},k}^2 q_{\text{D},k} - \gamma_{\rm v} q_{\rm B}  .\label{eq:EOM_sub_D}
        \end{align}
    \end{subequations}
     Here, for the electronic dynamics in Eq. \eqref{eq:EOM_sub_e-nodamping}, 
    $\hat{\mathcal{L}}_{\rm dmp}[\hat{\rho}_{\rm e,s}]$ denotes the electronic depopulation and decoherence Lindbladian:
$\hat{\mathcal{L}}_{\rm dmp}[\hat{\rho}_{\rm e,s}] \equiv \gamma_{\rm e} \left ( 
            \hsigma_{+}\hrho_{\rm e, s}\hsigma_{-} - \frac{1}{2}\hsigma_{+}\hsigma_{-}\hrho_{\rm e, s} - \frac{1}{2}\hrho_{\rm e, s} \hsigma_{+}\hsigma_{-}
            \right ) $ with the damping coefficient denoted as $\gamma_{\rm e}$; $\hat{\mathcal{L}}_{\rm drv}[\hat{\rho}_{\rm e,s}] = d_{\rm eg}E(t) ( \hat{\sigma}_{+} + \hat{\sigma}_{-} )$ is the additional external driving due to the coupling with the external pulse defined by $E(t)$. Because this pulse will be operated in the electronic frequency domain, it is assumed to be decoupled from the vibrational and the photonic degrees of freedom. 
    In Eq. \eqref{eq:EOM_sub_ph}, the $- \gamma_{\rm c} p_{\rm c}$ term is added phenomenologically to account for the  cavity loss due to the imperfectness of the cavity mirrors, where $\gamma_{\rm c}$ denotes the cavity loss rate.
    In Eqs. \eqref{eq:EOM_sub_B} and \eqref{eq:EOM_sub_D}, the terms containing $\gamma_{\rm v}$ are introduced to describe the polariton dephasing to the dark modes, where $\gamma_{\rm v}$ characterizes the coupling strength between the bright and the dark states. If $\gamma_{\rm v} = 0$, the bright state would be completely decoupled from the dark modes. For the dark-mode dynamics, because in experiments the dissipation lifetime of vibrational dark modes to the ground state can take hundreds of ps to even tens of ns (which is much longer than the timescale we are interested in) \cite{Xiang2019State}, here the dissipation of vibrational dark modes is not considered. In Eq. \eqref{eq:EOM_sub_D}, the vibrational frequencies of different dark modes are $\omega_{\text{D},k}$. The dark-mode frequencies will be sampled from a uniform distribution around $\omega_{\rm v}$ (the bright-mode frequency). Such a frequency distribution \cite{Nitzan2006} will allow an efficient simulation of the dissipation from the bight to the dark modes with an affordable number of dark modes. 

    The following set of parameters was used by default. For the electronic subsystem, $N_{\rm e}  = 10^{10}$, $\omega_{\rm e} = 0.1$ a.u. = 2.7 eV, $d_{\rm eg} = 0.5$ a.u., $d_{\rm gg} = 0$, $d_{\rm ee} = 1.0$ a.u., and $\gamma_{\rm e}$ = $10^{-5}$ a.u. ($1/\gamma_{\rm e} = 2.4$ ps).
    For the photon mode, $\omega_{\rm c} = 0.01$ a.u. = 2195 cm$^{-1}$, 
    $\lambda_{\rm c} = 2\times 10^{-6}$ a.u., and $\gamma_{\rm c} = 2\times 10^{-5}$ a.u. ($1/\gamma_{\rm c} = 1.2$ ps).
    For the vibrational subsystem,  $N_{\rm v}  = 10^{10}$, $\omega_{\rm v} = 0.01$ a.u. = 2195 cm$^{-1}$, and $d_{\rm v} = 0.01$ a.u.; for the dark states, instead of directly simulating $N_{\rm v}-1$ dark states directly, we used $N_{\rm dark} = 500$ harmonic oscillators with a uniform frequency distribution between 0.007 and 0.013 a.u., and the coupling strength between the bright and the dark states was set as $\gamma_{\rm v} = 2\times 10^{-6} /\sqrt{N_{\rm dark}}$ a.u.    For the external pulse, $E(t)= E_0\sin(\omega_{\rm e} t) \exp[-(t-t_{\rm start})^2/\sigma^2]$, where $E_0 = 0.01$ a.u., $t_{\rm start} = 5\times 10^{2}$ a.u. = 12.1 fs, and $\sigma = 100$ a.u. = 2.4 fs. In the manuscript, when the simulation results were scanned against one parameter, all other parameters were kept the same as the default values. Note that the simulation results have already converged when $N_{\rm dark} = 500$, and further increasing $N_{\rm dark}$ would not lead to any meaningful change of the data, in agreement with our previous simulations of polariton relaxation \cite{Li2020Nonlinear}. After all, simulating polariton relaxation to the dark modes alone does not require a macroscopically large number of molecules. Additionally, setting the coupling $\gamma_{\rm v} = 2\times 10^{-6} /\sqrt{N_{\rm dark}}$ a.u., i.e., an inverse dependence on $\sqrt{N_{\rm dark}}$, was to ensure that the polariton dephasing lifetime did not change when only $N_{\rm dark}$ was tuned.

    \begin{figure}
		\centering
		\includegraphics[width=0.5\linewidth]{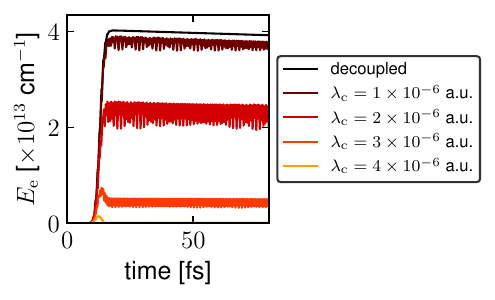}
		\caption{Electronic energy dynamics of $N_{\rm e}$ TLSs during an Gaussian pulse excitation for the coupled electron-vibration-cavity system. Lines with different colors represent simulations with varied cavity-matter coupling strength $\lambda_{\rm c}$ from zero (black) to $4\times 10^{-6}$ a.u. Increasing $\lambda_{\rm c}$ leads to suppressed electronic excitations. 
 		}
		\label{fig:traj_compare}
    \end{figure}

    \section{The inversion regimes in Figs. 3a,b}

    The inversion regimes in Figs. 3a,b, which are not predicted in Eqs. (7) and (8), are worthy of further investigation. For a better understanding of the inversion regimes, Fig. \ref{fig:traj_compare} plots the electronic energy dynamics of all TLSs under different cavity-matter coupling strengths $\lambda_{\rm c}$. Very interestingly, increasing  $\lambda_{\rm c}$ causes a suppression of the electronic energy gain under the same Gaussian pulse excitation. This electron-energy suppression at large $\lambda_{\rm c}$ values were not considered during the derivation of Eqs. (7) and (8), which assumed that the electronic TLSs were excited equally if the pulse excitation was the same, in spite of the change of any other parameters. If the electron-energy suppression under large $\lambda_{\rm c}$ values were considered, Eqs. (7) and (8) would have suggested that the cavity or vibrational energy gain becomes less efficient when  $\lambda_{\rm c}$ becomes very large. 

    The electron-energy suppression can be understood as follows.
    As shown in Eq. \eqref{eq:He_semi_single}, the terms containing $\lambda_{\rm c}$ or $N_{\rm v}$ can renormalize the electronic transition frequency $\omega_{\rm e}$, in a manner similarly as the Lamb shift outside the cavity. When $\lambda_{\rm c}$ or $N_{\rm v}$ is increased, the effective electronic transition frequency will be tuned away from $\omega_{\rm e}$. Because the external pulse frequency is always centered at $\omega_{\rm e}$, increasing $\lambda_{\rm c}$ or $N_{\rm v}$ would reduce the excitation of the electronic system, thus suppressing the supervibronic transition process.

    \section{Influence of relaxation pathways}

    \begin{figure}
		\centering
		\includegraphics[width=0.5\linewidth]{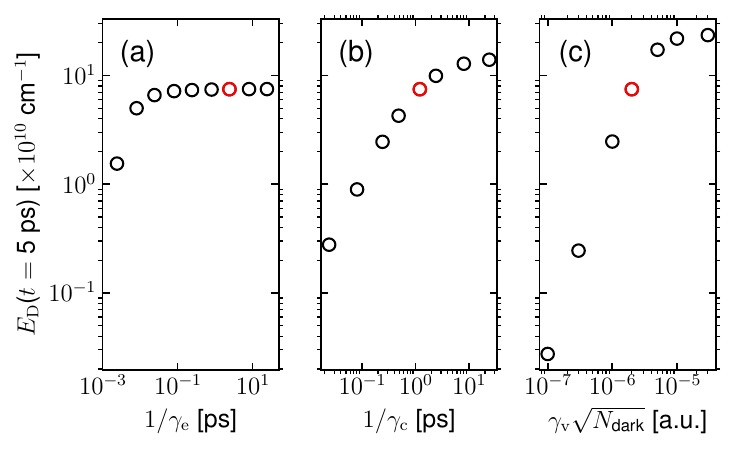}
		\caption{$E_{\rm D}(t = 5 \text{ ps})$ against different relaxation parameters in the logarithmic scale: (a) electronic relaxation lifetime $1/\gamma_{\rm e}$; (b) cavity loss lifetime $1/\gamma_{\rm c}$; (c) coupling strength between the vibrational bright and the dark modes $\gamma_{\rm v}$. The red circle in each subplot represents the data point corresponding to the simulation in Figs. 2e,f with the default parameters.
		}
		\label{fig:params_dephasing}
    \end{figure}

    Finally, Fig. \ref{fig:params_dephasing} plots the $E_{\rm D}(t = 5 \text{ ps})$  dependence on the three dissipation channels: the electronic relaxation rate $\gamma_{\rm e}$, the cavity loss rate $\gamma_{\rm c}$, as well as the coupling strength between the vibrational bright and dark modes $\gamma_{\rm v}$. Clearly, $E_{\rm D}(t = 5 \text{ ps})$ can range from $10^{10}$ to $10^{11}$ cm$^{-1}$ for a very wide range of dissipative rates. Because $N_{\rm v} = 10^{10}$ has been taken as the simulation value, it suggests that each vibrational harmonic oscillator could be excited by $1\sim 10$ cm$^{-1}$. This value indicates that the supervibronic transition mechanism may potentially be experimentally verified at room temperature, although at lower temperatures this mechanism would be more easily identified.

    \red{
    
    \section{More realistic simulations of the electronic subsystem}

    In the manuscript, we have restricted ourselves to the Dicke's superradiant states, i.e., assuming that the electronic dynamics of $N_{\rm e}$ TLSs can be propagated using the density matrix of a single TLS. This assumption can break down if the electronic TLSs experience different local environments, e.g., if the dipole orientations of the TLSs are not uniform. To examine this critical assumption, below we consider two approaches to evaluate the electronic dynamics of the system more realistically.

    \subsection{Direct propagation of the full electronic density matrix}

    For the electronic dynamics, instead of propagating Eq. \eqref{eq:EOM_sub_e}, we can directly propagate the density matrix for the whole electronic subsystem $\hrho_{\rm e}$ using:
    \begin{equation}\label{eq:EOM_e_full_dm}
        \frac{\partial}{\partial t} \hrho_{\rm e} = -i \left [ \hH_{\rm sc},\ \hrho_{\rm e} \right ] + \hat{\mathcal{L}}_{\rm dmp}[\hat{\rho}_{\rm e}] + \hat{\mathcal{L}}_{\rm drv}[\hat{\rho}_{\rm e}] .
    \end{equation}
    Here, the electronic Hamiltonian $\hH_{\rm sc}$ has been defined in Eq. \eqref{eq:Hsc}:
    \begin{equation*}
        \hH_{\rm sc} = \omega_{\rm e} \sum_{i=1}^{N_{\rm e}} \hat{\sigma}_{+}^{i}\hat{\sigma}_{-}^{i} + \omega_{\rm c}\lambda_{\rm c}q_{\rm c} \hat{\mu} + \frac{1}{2}\lambda_{\rm c}^2 \hat{\mu}^2 ,
    \end{equation*}
    where we have defined the semiclassical total dipole operator as
    \begin{equation}
        \hat{\mu} = \mu_{\rm v} + \hat{\mu}_{\rm e} = \sum_{j=1}^{N_{\rm v}} d_{\rm v} q_j + \sum_{i=1}^{N_{\rm e}} \cos\theta_i \left [  d_{\rm eg}( \hat{\sigma}_{+}^{i} + \hat{\sigma}_{-}^{i} ) + \left( \bar{d} \hat{1}^i - \frac{\Delta d}{2} \hat{\sigma}_z^{i} \right) \right ].
    \end{equation}
    Here, $\theta_i$ denotes the angle between the electronic dipole orientation of each TLS and the cavity polarization direction. If the dipole orientations of all the TLSs align with the cavity polarization direction uniformly, $\theta_i \equiv 0$  for $i = 1,2\cdots, N_{\rm e}$. For simplicity, the transition dipole moments of $N_{\rm v}$ vibrational harmonic oscillators  have been assumed to align with the cavity polarization direction uniformly (for now). The expectation value of the total dipole moment is 
    \begin{equation}
        \avg{\hat{\mu}} = \mu_{\rm v} + \tr{\hat{\rho}_{\rm e} \hat{\mu}_{\rm e}}.
    \end{equation}
    In Eq. \eqref{eq:EOM_e_full_dm}, $\hat{\mathcal{L}}_{\rm dmp}[\hat{\rho}_{\rm e}]$ denotes the electronic depopulation and decoherence Lindbladian for $N_{\rm e}$ TLSs: $\hat{\mathcal{L}}_{\rm dmp}[\hat{\rho}_{\rm e}] \equiv \sum_{i=1}^{N_{\rm e}}\gamma_{\rm e} \left ( \hsigma_{+}^i\hrho_{\rm e}\hsigma_{-}^i - \frac{1}{2}\hsigma_{+}^i\hsigma_{-}^i\hrho_{\rm e} - \frac{1}{2}\hrho_{\rm e} \hsigma_{+}^i\hsigma_{-}^i \right ) $ with the damping coefficient $\gamma_{\rm e}$; $\hat{\mathcal{L}}_{\rm drv}[\hat{\rho}_{\rm e}] = \sum_{i=1}^{N_{\rm e}}  d_{\rm eg}E(t) \cos\theta_i ( \hat{\sigma}_{+}^i + \hat{\sigma}_{-}^i ) $ is the additional external driving due to the coupling with the external pulse defined by $E(t)$. Because the dipole orientations of the TLSs have been taken into account, the prefactor $\cos\theta_i$ appears also in $\hat{\mathcal{L}}_{\rm drv}[\hat{\rho}_{\rm e}]$.

    At the beginning of the dynamics, the electronic density matrix is assumed to be the matrix product of the density matrices of all the TLSs:
    \begin{equation}
        \hat{\rho}_{\rm e}(t=0) = \prod_{i=1}^{N_{\rm e}} \hat{\rho}_{\rm e}^{i}(t=0) ,
    \end{equation}
    where $\hat{\rho}_{\rm e}^{i}(t=0) = \ket{gi}\bra{gi}$, i.e., all the TLSs start in their electronic ground states. Because the full electronic density matrix of size $2^{N_{\rm e}}\times 2^{N_{\rm e}}$ is propagated, during the simulation, quantum entanglement between different TLSs can be naturally captured. 
    
    \subsection{The assumption of matrix product state}

    The above brute-force simulation of the electronic TLSs is computationally very demanding, as the dimension of the electronic Hilbert space grows as $2^{N_{\rm e}}$. A less expensive propagation of the electronic TLSs can be obtained by assuming that the electronic density matrix takes the following matrix product form during the dynamics:
    \begin{equation}
        \hat{\rho}_{\rm e}(t) = \prod_{i=1}^{N_{\rm e}} \hat{\rho}_{\rm e}^{i}(t) .
    \end{equation}
    With this matrix-product-state approximation, the electronic dynamics can be described by $N_{\rm e}$  $2\times 2$ density matrices $\hrho_{\rm e}^i$ (for $i=1,2,\cdots,N_{\rm e}$) obeying:
    \begin{equation}\label{eq:EOM_e_dm_mp}
        \frac{\partial}{\partial t} \hrho_{\rm e}^i = -i \left [ \hH_{\rm sc}^i,\ \hrho_{\rm e}^i \right ] + \hat{\mathcal{L}}_{\rm dmp}[\hat{\rho}_{\rm e}^i] + \hat{\mathcal{L}}_{\rm drv}[\hat{\rho}_{\rm e}^i] .
    \end{equation}
    Here, the single-body semiclassical Hamiltonian $\hH_{\rm sc}^i$ reads
    \begin{equation}
        \hH_{\rm sc}^i = \omega_{\rm e} \hat{\sigma}_{+}^i \hat{\sigma}_{-}^i  + \omega_{\rm c}\lambda_{\rm c}q_{\rm c} \hat{\mu}_{\rm e}^i + \frac{1}{2}\lambda_{\rm c}^2 \left [(\hat{\mu}_{\rm e}^i)^2 + 2\sum_{i'\neq i}\avg{\hat{\mu}_{\rm e}^{i'}}\hat{\mu}_{\rm e}^i + 2\mu_{\rm v}\hat{\mu}_{\rm e}^i\right ] ,
    \end{equation}
    where the single-body electronic dipole operator is
    \begin{equation}
        \hat{\mu}_{\rm e}^i =  \cos\theta_i \left [  d_{\rm eg}( \hat{\sigma}_{+}^{i} + \hat{\sigma}_{-}^{i} ) + \left( \bar{d} \hat{1}^i - \frac{\Delta d}{2} \hat{\sigma}_z^{i} \right) \right ],
    \end{equation}
    and $\avg{\hat{\mu}_{\rm e}^{i}} = \tr{\hat{\rho}_{\rm e}^{i} \hat{\mu}_{\rm e}^{i} }$.
    The expectation value of the total dipole moment is 
    \begin{equation}
        \avg{\hat{\mu}} = \mu_{\rm v} + \sum_{i=1}^{N_{\rm e}}\tr{\hat{\rho}_{\rm e}^i \hat{\mu}_{\rm e}^i}.
    \end{equation}
    In Eq. \eqref{eq:EOM_e_dm_mp}, $\hat{\mathcal{L}}_{\rm dmp}[\hat{\rho}_{\rm e}^i]$ denotes the electronic depopulation and decoherence Lindbladian for the $i$-th TLS: $\hat{\mathcal{L}}_{\rm dmp}[\hat{\rho}_{\rm e}^i] \equiv \gamma_{\rm e} \left ( \hsigma_{+}^i\hrho_{\rm e}\hsigma_{-}^i - \frac{1}{2}\hsigma_{+}^i\hsigma_{-}^i\hrho_{\rm e} - \frac{1}{2}\hrho_{\rm e} \hsigma_{+}^i\hsigma_{-}^i \right ) $; the external driving is represented by $\hat{\mathcal{L}}_{\rm drv}[\hat{\rho}_{\rm e}] =  d_{\rm eg}E(t) \cos\theta_i ( \hat{\sigma}_{+}^i + \hat{\sigma}_{-}^i ) $.
    
    \subsection{Comparison of the three approaches on propagating electronic dynamics}

    \begin{figure}
		\centering
		\includegraphics[width=0.5\linewidth]{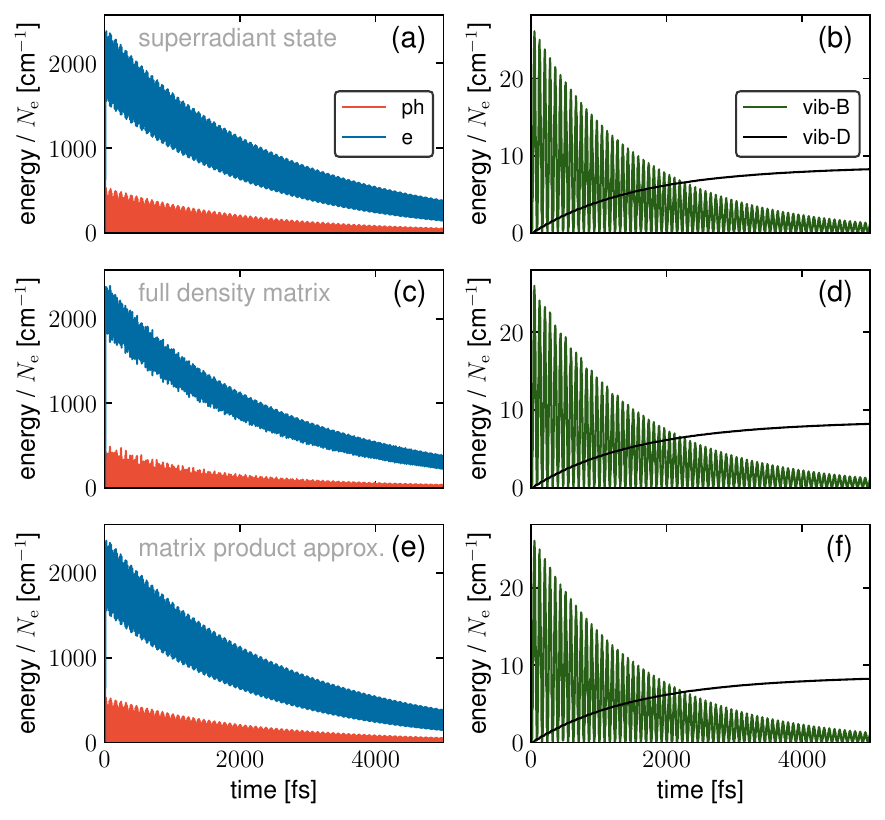}
		\caption{\red{Energy dynamics of different subsystems divided by the number of electronic TLSs. The electronic subsystem is propagated by (a,b) the density matrix of a single TLS; (c,d) a $2^{N_{\rm e}} \times 2^{N_{\rm e}}$ full density matrix; (e,f) $N_{\rm e}$ $2\times 2$ density matrices. For parameters, $N_{\rm e}=N_{\rm v}=4$ and the light-matter coupling $\lambda_{\rm c} = 0.1$ a.u. is unphysically large; all the other parameters remain the same as the default setting. The three approaches yield the same results.}
		}
		\label{fig:compare_three_edynamics}
    \end{figure}
    
    Assuming uniform dipole orientations for all the TLSs (i.e., $\theta_i\equiv 0$), we can directly compare the three approaches above for propagating the electronic dynamics: Dicke's superradiant states (i.e., using the density matrix of a single TLS), the full density matrix, and the matrix product state (i.e., using $N_{\rm e}$ $2\times 2$ density matrices). The photonic and vibrational degrees of freedom are still governed by the equations of motion in Eq. \eqref{eq:EOM_final_original}. 
    
    However, due to the high computational cost of simulating the full density matrix, the three approaches can only be compared together when $N_{\rm e}$ is very small. Hence, we  set  $N_{\rm e} = N_{\rm v} = 4$. Because the molecular number is very small, to reproduce similar dynamics as in the manuscript, we have to choose a unphysically large light-matter coupling strength: $\lambda_{\rm c} = 0.1$ a.u. by rescaling the default value of $\lambda_{\rm c}$: $\lambda_{\rm c} = \lambda_{\rm c}^{\rm default} \sqrt{N_{\rm e}^{\rm default} / N_{\rm e}}$. With this rescaling, the Rabi splitting remained the same.   All the other parameters were  the same as the default setting.

    Fig. \ref{fig:compare_three_edynamics} plots the energy dynamics of different degrees of freedom using the three different approaches for propagating the electronic dynamics: (a,b) using the Dicke's superradiant states (i.e., the single $2\times 2$ density matrix calculation); (c,d) using the full $2^{N_{\rm e}}\times 2^{N_{\rm e}}$ density matrix; and (e,f) using the matrix product state (i.e., using $N_{\rm e}$ $2\times 2$ density matrices). Their agreement on the dynamics validates the use of the Dicke's superradiant states in the manuscript, where the electronic TLSs are oriented uniformly and all the TLSs starts from the ground state.

    \subsection{Effects of electronic dipole orientations}

    \begin{figure}
		\centering
		\includegraphics[width=0.5\linewidth]{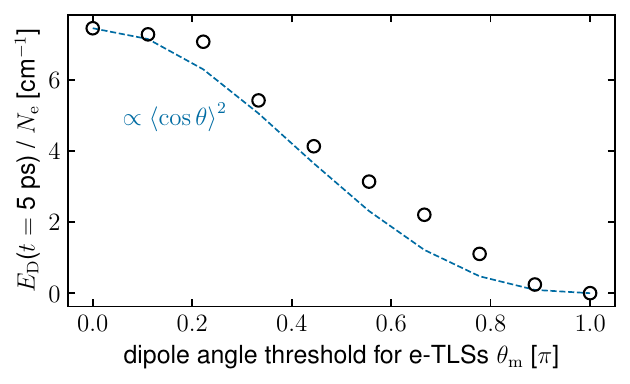}
		\caption{\red{Long-time energy gain of vibrational dark modes divided by the number of electronic TLSs [$E_{\rm D}(t=5 {\rm{\ ps}}) / N_{\rm e}$] as a function of the electronic dipole angle threshold $\theta_{\rm m}$. The electronic dipole angles (with respect to the cavity polarization direction) are distributed uniformly in the range of $[-\theta_{\rm m}, \theta_{\rm m}]$. The electronic subsystem is propagated with $N_{\rm e}$ $2\times 2$ density matrices. For parameters, $N_{\rm e}=N_{\rm v}=100$ and the light-matter coupling $\lambda_{\rm c} = 0.02$ a.u. is unphysically large; all the other parameters remain the same as the default setting. The results scale with $\avg{\cos\theta}^2$ (dashed blue line).}
		}
		\label{fig:params_dependence_eangle}
    \end{figure}

    To study the effects of electronic dipole orientations on the dynamics, we use the matrix product state approach (i.e., using $N_{\rm e}$ $2\times 2$ density matrices). In this calculation, we use a set of parameters that are nonphysical: $N_{\rm e} =  N_{\rm v} = 100$, and the light-matter coupling strength is $\lambda_{\rm c} = 0.02$ a.u. Similar to the case discussed above, here the coupling strength is rescaled as $\lambda_{\rm c} = \lambda_{\rm c}^{\rm default} \sqrt{N_{\rm e}^{\rm default} / N_{\rm e}}$ to ensure similar overall dynamics to those in the manuscript.   The electronic dipole angles $\theta_i$ obey a uniform distribution in the range of $[-\theta_{\rm m}, \theta_{\rm m}]$, where $\theta_{\rm m}$ is the threshold angle of the dipole orientation disorder. If the electronic dipole orientations are completely random, $\theta_{\rm m} = \pi$; if the electronic dipole orientations are uniform, $\theta_{\rm m} = 0$. All the other parameters remain the same as the default setting.

    Fig. \ref{fig:params_dependence_eangle} plots $E_{\rm D}(t=5 {\rm{\ ps}}) / N_{\rm e}$, the long-time energy gain of vibrational dark modes divided by the number of electronic TLSs, as a function of the electronic dipole angle threshold $\theta_{\rm m}$.  The simulation data scale with $\avg{\cos\theta}^2$ (the dashed blue line), indicating that the supervibronic transition mechanism disappears if the electronic dipoles have random orientations. However, if the electronic dipole orientations are anisotropic ($\avg{\cos\theta}^2\neq 0$), the supervibronic transition mechanism can still be preserved. The $\avg{\cos\theta}^2$ scaling is not surprising, as the supervibronic transition mechanism scales quadratically with the permanent dipole change in the electronic subsystem: $(\sum_{i=1}^{N_{\rm e}} \cos\theta \Delta d)^2 \propto \avg{\cos\theta}^2$.

    \section{Effects of vibrational dipole orientations}

    \begin{figure}
		\centering
		\includegraphics[width=0.5\linewidth]{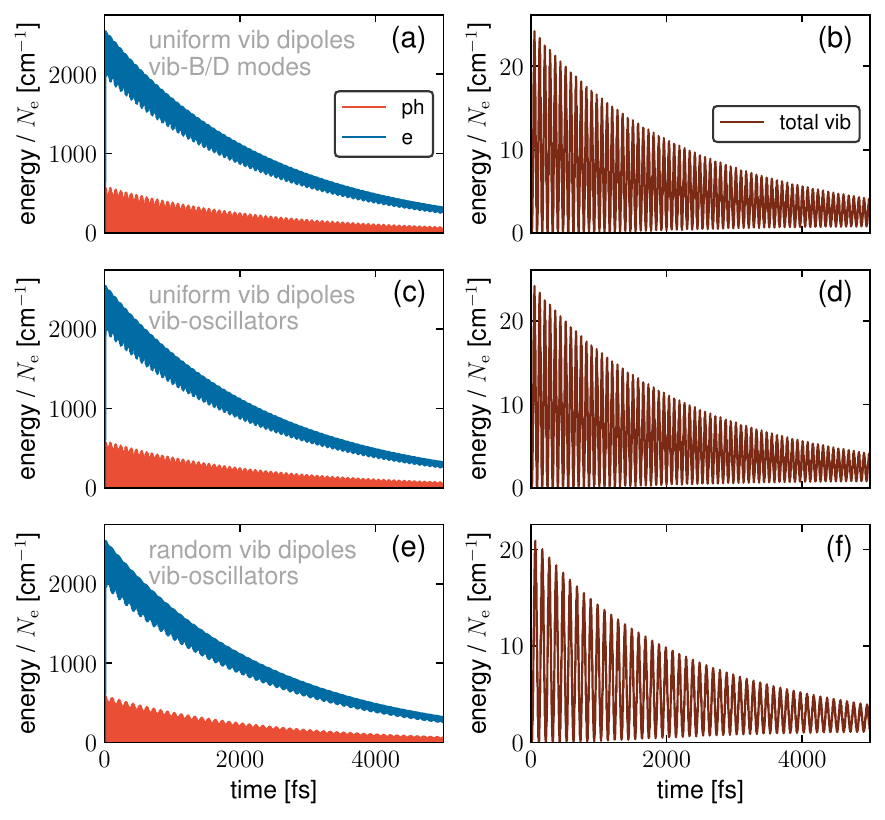}
		\caption{\red{Energy dynamics of different subsystems divided by the number of electronic TLSs. Energy of the photonic (red), electronic (blue), and vibrational (brown, bright + dark modes) degrees of freedom is shown under different conditions: (a,b) simulating the bright and the dark modes with uniformly distributed vibrational dipoles; directly simulating $N_{\rm v}$ vibrational harmonic oscillators with (c,d) uniformly and (e,f) randomly distributed vibrational dipoles. For parameters, $N_{\rm e}=N_{\rm v}=500$ and the light-matter coupling $\lambda_{\rm c} = 8.935\times 10^{-3}$  a.u. is unphysically large. Due to the difficulty of simulating the polariton dephasing  to the dark modes when propagating individual vibrational harmonic oscillators, this dephasing channel is turned off for all the calculations. All the other parameters remain the same as the default setting. Clearly, with a random distribution of the vibrational dipole orientations, the Rabi oscillation period is increased, but the magnitude of energy transfer to the vibrational degrees of freedom is not changed.}
		}
		\label{fig:compare_vib_orientation}
    \end{figure}

    After quantifying the effects of electronic dipole orientations on the dynamics, we are also interested in the role of vibrational dipole orientations. Because the equations of motion in Eq. \eqref{eq:EOM_final_original} propagate the collective vibrational bright and dark modes, this  approach cannot account for the explicit role of the vibrational dipole orientations.

    For a explicit simulation of the vibrational dipole orientations, the vibrational dynamics of each vibrational harmonic oscillator can be propagated directly:
    \begin{equation}\label{eq:EOM_vib_osc}
        \ddot{q}_{j} = -\omega_{\rm v}^2 q_{j} - \omega_{\rm c}\lambda_{\rm c} q_{\rm c} \frac{\partial \avg{\hat{\mu}}}{\partial q_{j}} 
            - \lambda_{\rm c}^2 \avg{\hat{\mu}} \frac{\partial \avg{\hat{\mu}}}{\partial q_{j}} .
    \end{equation}
    Assuming that the electronic and photonic dynamics still obey Eq. \eqref{eq:EOM_final_original} (and the electronic dipole orientations are still uniform), we can calculate the expectation value of the total dipole moment with  $\avg{\hat{\mu}} = \sum_{j=1}^{N_{\rm v}}\cos\alpha_j d_{\rm v} q_{j} + N_{\rm e}\tr{\hrho_{\rm e,s}\hat{\mu}_{e,s}}$, where $\alpha_j$ denotes the angle between each vibrational transition dipole orientation and the cavity polarization direction. The dipole derivative can then be evaluated as $\partial \avg{\hat{\mu}} / \partial q_{j} = \cos\alpha_j d_{\rm v}$.

    The disadvantage of using Eq. \eqref{eq:EOM_vib_osc} is that it cannot easily describe the mechanism of vibrational polariton dephasing to the dark modes  under the basis of individual sites. In contrast, with the basis of the collective bright and dark modes, the dephasing can be described with a simple coupling term between the bright and the dark modes [see Eq. \eqref{eq:EOM_final_original}]. With the basis of individual vibrational harmonic oscillators, however, additional disorder or the coupling to independent thermal baths \cite{Pino2015} may be necessary  to recover the dephasing. Due to the complexity of considering these factors, we now compare the results obtained from Eq. \eqref{eq:EOM_vib_osc} with those from Eqs. \eqref{eq:EOM_sub_B} and \eqref{eq:EOM_sub_D} by turning off this dephasing channel. In other words, with the basis of independent vibrational harmonic oscillators, Eq. \eqref{eq:EOM_vib_osc} is directly propagated, whereas with the basis of the vibrational bright and the dark modes, Eqs. \eqref{eq:EOM_sub_B} and \eqref{eq:EOM_sub_D} are propagated with the coupling between the bright and the dark modes turned off ($\gamma_{\rm v} = 0$).

    Fig. \ref{fig:compare_vib_orientation} depicts the energy dynamics of different degrees of freedom under three conditions:  uniformly distributed vibrational dipole orientations  (a,b) when the bright and the dark modes are simulated [Eqs. \eqref{eq:EOM_sub_B} and \eqref{eq:EOM_sub_D}] or (c,d) when individual vibrational harmonic oscillators are simulated [\eqref{eq:EOM_vib_osc}]; (e,f) randomly distributed vibrational dipole orientations when individual vibrational harmonic oscillators are simulated [\eqref{eq:EOM_vib_osc}]. Here, in the case where the vibrational dipole orientations are random, the amount of energy received by the vibrational degrees of freedom is not altered. However, the Rabi oscillation period is shortened when the dipole orientations become random. Hence, the vibrational dipole orientations do not significantly alter the energy transfer.

    This behavior can be understood as follows. The excitation of the Casimir IR photons, according to Eq. (7) in the manuscript, is independent of the vibrational dipole orientations. Hence, the energy received by the vibrational degrees of freedom is independent of these orientations.  The orientations of the vibrational dipole would only change the coupling to the IR photon mode, or the Rabi splitting.

    \section{The importance of timescale separation and the permanent dipole difference $\Delta d$}

    \begin{figure}
		\centering
		\includegraphics[width=1.0\linewidth]{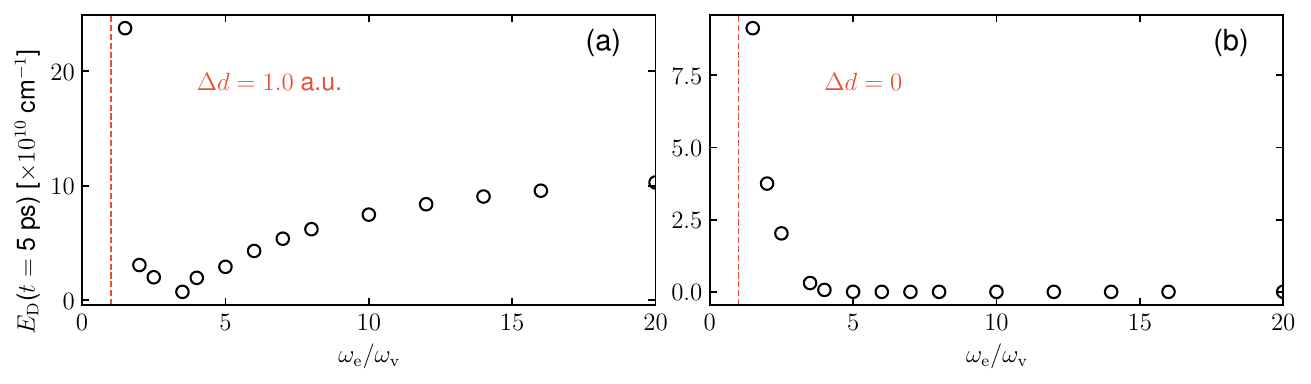}
		\caption{\red{Long-time vibrational energy gain [$E_{\rm D}(t=5 {\rm{\ ps}})$] versus the electronic transition frequency $\omega_{\rm e}$ by scanning different values of $\omega_{\rm e}$. Two conditions are considered: the permanent dipole difference between the electronic ground and the excited state (a) $\Delta d=1.0$ a.u. and (b) $\Delta d=0$. The equations of motion in Eq. \eqref{eq:EOM_final_original} are propagated, and the other parameters remain as the default setting. The dashed red lines represent $\omega_{\rm e} = \omega_{\rm v}$. For both cases, when $\omega_{\rm e}$ approaches $\omega_{\rm v}=\omega_{\rm c}$, resonance conditions permit the energy transfer. Only when $\Delta d\neq 0$, the supervibronic transition process occurs when $\omega_{\rm e}\gg \omega_{\rm v}, \omega_{\rm c}$.}
		}
		\label{fig:params_dependence_omegae}
    \end{figure}

    In contrast to conventional energy transfer mechanisms which require a resonance condition between the initial and final states, the supervibronic transition mechanism reported in the manuscript requires a timescale separation between the electronic and vibrational (or photonic) degrees of freedom.

    Fig. \ref{fig:params_dependence_omegae}a plots the long-time vibrational energy gain [$E_{\rm D}(t=5 {\rm{\ ps}})$] as a function of  the electronic transition frequency $\omega_{\rm e}$ when $\Delta d = 1.0$ a.u. In two parameter limits vibronic transitions occur: near resonance ($\omega_{\rm e}/\omega_{\rm v}\rightarrow 1$, the dashed red line) and when the timescale separation becomes valid ($\omega_{\rm e}/\omega_{\rm v} > 5$). However, as shown in Fig. \ref{fig:params_dependence_omegae}b, when the electronic permanent dipole difference $\Delta d = 0$, the supervibronic transition process no longer occur  when $\omega_{\rm e}/\omega_{\rm v} \gg 1$. Overall, Fig. \ref{fig:params_dependence_omegae} highlights the important roles of  both the timescale separation and the permanent dipole difference $\Delta d$ on facilitating the supervibronic transition mechanism.

    }
    
	
	%